\documentstyle[12pt]{article}

0

\begin{document}

\newcommand{\MqV}{$\cal M$$_{q}$($\cal V$)}
\def\II{\relax{\rm 1\kern-.35em1}}
\def\IP{\relax{\rm I\kern-.18em P}}
\renewcommand{\theequation}{\thesection.\arabic{equation}}
\csname @addtoreset\endcsname{equation}{section}

\begin{flushright}
hep-th/9510023
FTUAM 95/36
\end{flushright}
\vglue 1cm

\begin{center}

{}~\vfill

{\large \bf  ELECTRIC-MAGNETIC DUALITY AND \protect \\
{}~\vfill
EFFECTIVE FIELD THEORIES$^{\dag}$}

\vspace{20 mm}

{\large C\'{e}sar G\'{o}mez}

\vspace{1cm}

{\em Instituto de Matem\'{a}ticas y F\'{\i}sica Fundamental, CSIC \protect \\
Serrano 123, 28006 Madrid, Spain}
\vspace{1cm}

and

\vspace{1cm}

{\large Rafael Hern\'{a}ndez}

\vspace{1cm}

{\em Departamento de F\'{\i}sica Te\'{o}rica, C-XI, Universidad Aut\'{o}noma de
Madrid \protect \\ Cantoblanco, 28049 Madrid, Spain}

\end{center}

\vspace{6 cm}

\dag  \hspace{2 mm}Lectures given by C.G. in the Advanced School on Effective
Theories,
Almu\~{n}ecar, Granada, 1995.

\pagebreak


\tableofcontents

\pagebreak


\section{Introduction.}

The very concept of law of nature reflects a terminology which appears to
be the heritage of a normative metaphor rooted in the ancestral image of a
universe ruled by God. Maybe, a deeper concept which underlies more basically
our way of thinking in physics is that of symmetry, which moves us from the
normative or legal metaphor to the belief that beauty is the closest to
truth. This was partially Dirac's philosophy of physics; these lectures will
try to present some of the recent and exciting developments in quantum field
theory and string theory that his original ideas of duality and magnetic
monopoles have made possible.

It was on the basis of the original idea of electric-magnetic duality
\cite{Dirac}
that the physics of magnetic monopoles begins \cite{GO}. The classical
mass formulas \cite{BPS} for the `t Hooft-Polyakov monopole \cite{tp}, together
with Witten's discovery of the dyon nature of the monopole and the role played
by the $\theta$-parameter \cite{t}, were crucial to promote the
electric-magnetic duality transformations into a group of symmetries, namely
the
modular group $Sl(2,Z)$. The unexpected connection between the BPS mass formula
and the $N\!=\!2$ supersymmetry algebra discovered by Olive and Witten
\cite{qm}
was already the first indication that a deep relation between duality and
supersymmetry existed \cite{o}.

In a different context, what might be called a ``dual'' way of thinking turned
to be an extremely powerful tool \cite{conf} to get a qualitative understanding
of the phenomena of confinement: we only need to think of the dual to the well
known
Meissner effect in BCS-superconductivity, to interpret confinement as the dual
to the Higgs mechanism.

However, the strong constraints impossed by the Montonen-Olive duality
conjecture \cite{dual} reduce at first sight its field of application to
phenomenologically
uninteresting theories with vanishing $\beta$-function (for instance,
$N\!=\!4$ supersymmetric Yang-Mills). It was only very recently, thanks to the
seminal work of Seiberg and Witten \cite{SW}, that we have learned how to
extend, in a fruitful way, the ideas of duality to more realistic theories,
with non vanishing $\beta$-function. Besides, Seiberg-Witten results are not
only interesting from the strict sense of duality, as they deal with a
phenomenon
specially important in the phenomenology of supersymmetric theories (as well as
in string theories): the question of the physics associated with the existence
of flat potentials. The natural tool for studying this physics is, of course,
effective lagrangians, and that is the main reason why these lectures will
appear in this volume. A flat potential forces us to deal, from the beginning,
with a manifold of unequivalent field theories; the geometry of this manifold
can be described by the effective field theories that result from integrating,
at each point in the flat direction, the heavy modes. Now, typical questions
in the effective lagrangian physics, as the range of validity of a particular
choice of light degrees of freedom, acquires a new geometrical meaning, namely
the structure of singularities of the quantum moduli.

All these recent developments in quantum field theory have become possible, in
part,
thanks to the new way of thinking physics provided by string theory. In fact,
Montonen-Olive duality was rediscovered in the context of string theory
(interpreted as two dimensional $\sigma$-model physics) in what is called
$T$-duality \cite{T}. In this framework, the dynamical origin of this
symmetry is located in the extended nature of the fundamental string.
Montonen-Olive duality refers, of course, to four dimensional physics, but the
natural way in which strings explain the two dimensional analog strongly forces
us to look for the meaning of duality in a stringy framework. This was the
idea for postulating a new stringy symmetry, that was christened $S$-duality
\cite{S}.
Recently \cite{mr}, many new results
indicate that such a dream is close to the truth; the discovery of the so
called duality of dualities provides a bridge to connect $T$ and $S$-duality
in pairs of string theories. These fascinating topics dealing with string
theory
will, however, only be briefly treated at the end of the lectures.

In summary, the concept of duality is becoming an impressively useful tool
to work in many areas of theoretical physics, ranging from confinement and
effective field theories to extended supersymmetry and string theory.


\section{The Duality Group.}
\subsection{The Dirac Monopole.}

Maxwell's equations, governing the behaviour of the electromagnetic field,
offer a compact form when written in relativistic notation; in the absence of
sources, they look like
\begin{eqnarray}
           \partial_{\nu}{\cal F}^{\mu \nu} & = & 0 \nonumber \\
\partial_{\nu} \, ^{\ast}{\cal F}^{\mu \nu} & = & 0
\label{eq:Mvac}
\end{eqnarray}
if a dual tensor of the electromagnetic tensor $\cal F$$_{\mu \nu}$ is
introduced:
\begin{equation}
^{\ast}{\cal F}^{\mu \nu} = \frac {1}{2} \epsilon ^{\mu \nu \rho \sigma}
{\cal F}_{\rho \sigma},
\end{equation}
with $\epsilon ^{\mu \nu \rho \sigma}$ the totally antisymmetric tensor.

These equations in vacua are obviously symmetric under the {\em duality }
transformation
\begin{equation}
{\cal F}^{\mu \nu} \rightarrow  ^{\ast}{\cal F}^{\mu \nu} \: \: \: \: \: \: \:
\: \: \: \: \:
^{\ast}{\cal F}^{\mu \nu} \rightarrow - {\cal F}^{\mu \nu}
\label{eq:Mvd}
\end{equation}
which amounts to interchanging the role played by
electricity and magnetism. This symmetry is immediately broken if a non zero
electric current $j^{\mu}$ enters the theory, unless a magnetic current
$k^{\mu}$ is introduced, leading Maxwell's equations to the form
\begin{eqnarray}
           \partial_{\nu} {\cal F}^{\mu \nu} & = & - j^{\mu} \nonumber \\
\partial_{\nu} \, ^{\ast} {\cal F}^{\mu \nu} & = & - k^{\mu}.
\label{eq:M}
\end{eqnarray}
Duality appears again under the transformation, generalizing (\ref{eq:Mvd}),
\begin{eqnarray}
{\cal F}^{\mu \nu}  \rightarrow  ^{\ast} {\cal F}^{\mu \nu} & \> \> &
j^{\mu}  \rightarrow   k^{\mu}  \nonumber \\
^{\ast} {\cal F}^{\mu \nu}  \rightarrow - {\cal F}^{\mu \nu} &
\> \> &
k^{\mu}  \rightarrow  - j^{\mu}.
\label{eq:Mdual}
\end{eqnarray}

The immediate step one should give is wondering about whether this duality is
consistent with quantum theory. Quantization relies on the power of the
canonical
formalism, so we should keep track of the canonical variables for the
electromagnetic field, which are not the components of the tensor
$\cal F$$^{\mu \nu}$, but of the potential $A^{\mu}$, defined through
\begin{equation}
{\cal F}^{\mu \nu} = \partial^{\mu} A^{\nu} - \partial^{\nu} A^{\mu}.
\label{eq:A}
\end{equation}
However, finding a four-vector satisfying (\ref{eq:A}) is only possible if the
magnetic (dual) equation in (\ref{eq:M}) vanishes:
\begin{equation}
\partial_{\nu} ^{\ast}{\cal F}^{\mu \nu} = 0.
\end{equation}
But this condition, implied by equation (\ref{eq:A}), seems to destroy the
duality
coming from equations (\ref{eq:M}). Fortunately, there is an approach able to
mantain the chance to build an electromagnetic potential; all we have to do
is notice that in the neighbourhood of a magnetic charge ( a monopole)
the electromagnetic potential must be singular.
To see this, let us suppose that somewhere in space we have
a magnetic monopole of charge $g$, implying the nonvanishing of the magnetic
current $k^{\mu}$. Then, the magnetic flux leaving a sphere surrounding the
monopole can be easily calculated if we suppose that the electromagnetic
potential
is nonsingular everywhere. If we decompose $S^{2}$ into the two hemispheres
$H^{+}$ and $H^{-}$, and use Stoke's theorem, we will have a contribution
\begin{equation}
\int \int _{H^{+}} {\cal F}_{\mu \nu} d \, \Sigma ^{\mu \nu} =
\oint _{\partial H^{+}} A_{\mu} d \, x^{\mu},
\end{equation}
where the line integral, along the equator (the boundary of $H^{+}$), is taken
in the clockwise direction. Integrating over $H^{-}$,
\begin{equation}
\int \int _{H^{-}} {\cal F}_{\mu \nu} d \, \Sigma ^{\mu \nu} =
\oint _{\partial H^{-}} A_{\mu} d \, x^{\mu},
\end{equation}
where now the line integration is in the counterclockwise direction. Adding the
two integrals implies a zero for the total magnetic flux. As this contradicts
the assumption that the surface contains a magnetic monopole, in which case the
flux must be $4 \pi g$, we conclude that $A_{\mu}$ must have a singularity
somewhere on the sphere.

The argument above can be used for any radius of the sphere surrounding
the monopole, so by increasing it from zero to infinity we conclude that the
monopole
has attached a line of singularities. Dirac \cite{Dirac} was the first to
notice
such a line,
which is known as the {\em Dirac string}. The magnetic charge introduced to
make Maxwell's equations appear symmetric, and giving rise to the line of
singularities described, is the magnetic partner of the electron, and is
called a {\em Dirac monopole}.

Therefore, in the presence of a monopole the electromagnetic potential can not
be defined everywhere; all what can be done is find it everywhere except on a
line
joining the monopole to infinity. The orientation of the string is, of course,
arbitrary, and potential configurations in which the singularity extends
along different lines are related by gauge transformations.

\vspace{5 mm}

The Dirac string should not be thought of as a physical singularity, but as a
singularity in the representation of the potential in a particular gauge
choice.
It has the same meaning as the singularities in a stereographic projection
of the sphere: one can not help the appearence of a singular coordinate, say
the north or south pole, unless two different coordinate patches are used
in the description; in this case there is, in addition, the need to specify
which point in one projection corresponds to which point in the other
projection in the overlap region of the projections of the two patches. The
singularity in the electromagnetic potential coming from the presence of a
monopole can, in just the same way, be understood as the need to use more
than one coordinate patch to describe the potential, a perfectly licit
manipulation if we use gauge symmetry transformations to pass from one patch
to another.

\begin{quote}
The formulation given by Wu and Yang \cite{WY} of the above fact leads to the
mathematical meaning of magnetic charges. In the mathematical approach to
monopoles, the sphere $S^{2}$ surrounding the monopole becomes the base
space of a $U(1)$ (we are dealing with electromagnetism, with gauge group
$U(1)$) principal fibre bundle. When a connection satisfying Maxwell's
equations is chosen, we get a description of a magnetic monopole; in local
coordinates, the connection $1$-form $\omega$ is written as
\begin{equation}
\omega = g^{-1} A g + g^{-1} d g
\end{equation}
where \(A(x)=A_{\mu}^{a}(x) \lambda_{a}/2 i \, dx^{\mu}\) is the potential
$1$-form, with $\lambda_{a}/2 i$ an element of the Lie algebra.
For a $U(1)$ principal bundle $g=e^{i \psi}$, so the connection is simply
\begin{equation}
\omega = A + d \psi.
\end{equation}

Dividing $S^{2}$ into $H^{+}$ and $H^{-}$,
\begin{equation}
\omega = \left\{ \begin{array}{ll}
                A^{+} + d \psi^{+}, & \mbox{on } H^{+} \nonumber \\
                A^{-} + d \psi^{-}, & \mbox{on } H^{-}
                \end{array}
         \right.
\end{equation}

The transition functions must depend on the coordinates describing the
overlap region $H^{+}\cap H^{-}$, so they must now be functions of $\varphi$;
besides, the transition must take place through elements of the gauge group, so
\begin{equation}
e^{i \Psi^{-}}=e^{i n \varphi}e^{i \Psi^{+}}.
\end{equation}
Equivalently, the potential $1$-forms in the two hemispheres are related by
\begin{equation}
A^{+}=A^{-}+n \, d \varphi,
\end{equation}
that is, $A_{\mu}^{+}$ and $A_{\mu}^{-}$ are related by a gauge transformation:
\begin{equation}
A_{\mu}^{+}=A_{\mu}^{-} + \partial _{\mu}\lambda.
\label{eq:A+}
\end{equation}

An important conclusion follows from equation (1.13): in order to make sure
that the resulting structure is a manifold, the fibers must fit together
exactly
when completing full revolutions around the equator, so
{\em n must be an integer \/}\footnote{For $n=0$, we have a trivial bundle:
$S^{2}\times S^{1}$ ($S^{1}\simeq U(1)$). When $n=1$, we have the Hopf fibering
of the sphere $S^{3}$, describing a Dirac monopole of charge one, so the Dirac
monopole is a non trivial $U(1)$ principal fibre bundle with base $S^{2}$.}.

The monopole charge can be shown to coincide with the integral of the first
Chern class $c_{1}$ for the Dirac monopole $U(1)$ bundle over $S^{2}$: the
first Chern class is, up to a $2 \pi$ factor, the curvature $2$-form
$\cal F$\footnote{To be precise, the curvature for a $U(1)$ principal bundle
is purely imaginary and can be written as $\Omega=i F$.}
(the electromagnetic field strength); in fact,
\begin{equation}
c_{1} = - \frac {{\cal F}}{2 \pi}
\end{equation}
Applying Stoke's theorem, with $F=dA$ in mind, and taking into account that $A$
is separetely defined in $H^{+}$ and $H^{-}$, we get
\begin{equation}
\int_{S^{2}} c_{1} = - \int_{S^{2}} \frac {{\cal F}}{2 \pi} =
- \frac {1}{2 \pi}[\int_{H^{+}} dA^{+} + \int_{H^{-}} dA^{-}] =
- \frac {1}{2 \pi} \int_{S^{1}} A^{+} - A^{-}
\end{equation}
{}From (1.14), we then get
\begin{equation}
\int_{S^{2}} c_{1} = - \frac {1}{2 \pi} \int_{S^{1}} n \, d \varphi = - n.
\label{eq:chern1}
\end{equation}
\end{quote}

Comparing now (\ref{eq:A+}) and (\ref{eq:chern1}), we observe that the magnetic
charge
of the monopole can be directly interpreted as the winding number of the gauge
transformation $\partial \lambda$, which defines a map from the overlap region,
the equator, to the gauge group $U(1)$:
\begin{equation}
\partial \lambda : S^{1} \rightarrow U(1) \sim S^{1}.
\end{equation}
These maps are classified (see below) by the first homotopy group,
$\Pi_{1}(U(1))
\sim Z$, with the corresponding integer number, given by (\ref{eq:chern1}),
representing the winding number of the map.

\vspace{5 mm}

In classical electrodynamics, the whole theory is described in terms of the
electromagnetic field tensor $\cal F$$^{\mu \nu}$; however, when entering
quantum theory the knowledge of $\cal F$$^{\mu \nu}$ does not allow us to
determine the phase of the electron wave function, as the
{\em Aharanov-Bohm effect\/} shows: the potential appears again as the proper
tool in quantum theory, as it bears that information. When paralelly
transporting
a wave function along a path $\Gamma$, it picks up a {\em Dirac phase factor\/}
\begin{equation}
\exp [ i e \int_{\Gamma} A_{\mu}(x) \, dx^{\mu}],
\end{equation}
where $A_{\mu}$ is the potential due to a monopole. Then,
\begin{equation}
\Psi(P) \rightarrow \Psi(P')= \exp [ie \int_{\Gamma} A_{\mu} \, dx^{\mu}] \,
\Psi(P).
\end{equation}
As for a closed trajectory (describing a closed trajectory with our electron
amounts to looping once around the Dirac string) we must have
\begin{equation}
\Psi(P)=\Psi(P'),
\end{equation}
the phase factor must become the identity, implying
\begin{equation}
eg=2 \pi n, \> \> n \in Z.
\end{equation}
This is {\em Dirac quantization condition\/}, and contains a deep consequence:
the existence of magnetic charge implies the quantization of electric charge.
There
are many other different ways to get it; a topological version of it is the
comment stated above following from equation (1.13).

{}From the Dirac quantization condition we notice an extremely interesting
result:
the interchange of the role played by electricity and magnetism (duality),
obtained by exchanging the coupling constants, implies the interchange of
strong and weak coupling.

It is important te stress that the Dirac magnetic monopole is not part of
the spectrum of standard QED. Moreover, we can not define a local field
theory possesing, as part of its physical spectrum, both electrons and Dirac
monopoles. In order to use Dirac's duality symmetry in a more fundamental
way, we should look for local field theories which contain, as a part of the
physical spectrum, both electrically and magnetically charged particles. As
it was discovered by `t Hooft and Polyakov, spontaneously broken non abelian
gauge theories satisfying some topological criteria on its vacuum manifold
posses classical field configurations, which are solutions to the equations of
motion, topologically stable, magnetically charged, and have particle
like behaviour.

\subsection{The `t Hooft-Polyakov Monopole.}
Let us consider the Georgi-Glashow model. It consists of an $SO(3)$ gauge
field,
interacting
with an isovector Higgs field $\phi$:
\begin{equation}
{\cal L}= - \frac {1}{4}{\cal F}_{a}^{\mu \nu} {\cal F}_{a \mu \nu} +
\frac {1}{2} {\cal D}^{\mu} \phi \cdot {\cal D}_{\mu} \phi - V(\phi),
\> a=1,2,3,
\label{eq:LGG}
\end{equation}
where ${\cal D}^{\mu}_{a}$ is the covariant derivative, and $V(\phi)$ the
Higgs potential:
\begin{equation}
V(\phi)= \frac {1}{4} \lambda (\phi^{2}-a^{2})^{2}.
\label{eq:Hpot}
\end{equation}
In a Yang-Mills-Higgs configuration (a configuration where there is a
gauge in which \( \partial_{0}A=
\partial_{0}\phi=0 \)) with no time component, $A_{0}=0$, the energy
can be written in the simple form
\begin{equation}
E=\int [\frac {1}{2} \mid {\cal F} \mid^{2} + \frac {1}{2} \mid {\cal D} \phi
\mid ^{2} + V(\phi)]\, d^{3}r.
\label{eq:energy}
\end{equation}

Independently, `t Hooft and Polyakov \cite{tp} realized that the Georgi-Glashow
model
contains some remarkable finite energy solutions. As the Higgs potential
appears in the energy density, the integral (\ref{eq:energy}) can only converge
if, for
large distances, the Higgs field tends to the constant value $a$; a
configuration satisfying this is
\begin{equation}
\phi_{a}=a \delta_{a3}.
\end{equation}
Besides, the energy is gauge invariant, so any gauge rotation of this
configuration is
also a finite energy solution.

The set of $\phi$ which minimizes $V(\phi)$ constitutes the vacuum; in this
case,
the {\em vacuum manifold\/} $\cal V$, as follows from equation
(\ref{eq:Hpot}), is a two-dimensional sphere of radius $a$. The structure of
this manifold is determined by the gauge group $G$, and a subgroup in it known
as the little group:

The little group $H$ consists of the elements of $G=SO(3)$ leaving a given
$\phi$ invariant, so it is the group of rotations around the $\phi$ axis
(of radius a), that is, $SO(2)$; equivalently, $H\simeq U(1)$. Therefore,
finite energy enforces the gauge group $G$ to break, at large distances, down
to $H=U(1)$; the only field component remaining massless is that associated
with the residual $U(1)$, component we identify with the photon.

Assuming that $G$ acts transitively on $\cal V$, that is, given two fields
in $\cal V$ there is an element in $G$ relating them (\(\phi_{1},\phi_{2}
\in {\cal V} \Rightarrow \exists g_{12},\phi_{1}=D(g_{12})\phi_{2},\) with
$D$ a representation of $G$), the structure of the vacuum $\cal V$ is
determined by $G$ and $H$: the vacuum is the space of right cosets of $H$
in $G$,
\begin{equation}
{\cal V}=G/H.
\end{equation}

\vspace{5 mm}
To understand the meaning of this finite energy solution, we must have a
closer look around the region where it is located: so far, we have
dealt with the asymptotic properties the solution
must satisfy. In the regions where the gauge symmetry is unbroken, determining
a solution to the field equations is a difficult task, unless some simplifying
ansatz is used; the one used by `t Hooft and Polyakov is a spherically
symmetric one. With this ansatz, the field strength behaves, at large
distances,
as
\begin{equation}
{\cal F}_{a}^{ij} \sim \frac {1}{aer^{3}} \epsilon_{ijk} r^{k} \phi_{a}.
\end{equation}
The only surviving component of this field is that associated with the
neutral vector boson, the photon, and yields a magnetic field
\begin{equation}
B^{i}=- \frac {1}{e} \frac {r^{i}}{r^{3}}.
\end{equation}
Using Dirac quantization condition (and the fact that the smallest charge
which might enter the theory is not $e$, but $\frac {1}{2}e$) we notice
that, at large distances, our solution behaves like one unit of magnetic
charge (that is, a monopole).

Therefore, the `t Hooft-Polyakov monopole, at short distances, mantains all
fields
excited, giving rise to an $SU(2)$ symmetric finite energy solution; at
large distances the non-abelian symmetry is broken and all fields but the
residual photon are unexcited, giving rise to a solution that resembles a
Dirac monopole of unit magnetic charge. The ansatz given by `t Hooft and
Polyakov allows to define a Compton wavelength, such that the monopole can
be thought of as having a definite size, of the order $1/m_{W}$; in the inside,
the massive fields
provide a smooth structure and, in the outside, they vanish, leaving a field
configuration indistinguishable from the Dirac monopole.

However, the smooth internal structure satisfying $SO(3)$ gauge theory
equations of the `t Hooft-Polyakov monopole implies that there is no need
to introduce string singularities, in contrast to the Dirac monopole.

\vspace{5 mm}

To summarize, the field configurations obtained by `t Hooft and Polyakov
satisfy:
\begin{itemize}
        \item They are finite energy solutions.
        \item They represent magnetically charged states.
\end{itemize}
The monopoles contain, besides, a rich property, that of {\em topological
stability\/}, as will be pointed in the next paragraph.

\subsubsection{Topological Properties of Monopoles.}
We should first notice that as for large distances $\phi$ does not depend on
$r$, it provides a mapping from the sphere at infinity into the vacuum
manifold $\cal V$ (which is also a two-sphere):
\begin{equation}
\phi:S^{2} \rightarrow {\cal V}.
\end{equation}
This map admits a {\em winding number\/}, an integer representing the number
of times $\phi$ covers the sphere $\cal V$ as $(\theta,\varphi)$ covers once
the sphere at infinity.

Another interesting remark, that of topological stability, comes from observing
that outside the monopole (that is, for a radius larger than its Compton
wavelength) the field configuration is close to a Higgs vacuum,
\begin{equation}
{\cal D}_{\mu}\phi=\partial_{\mu}\phi - e A_{\mu}\wedge \phi = 0.
\label{eq:Hvac}
\end{equation}
The form of the potential satisfying (\ref{eq:Hvac}) is
\begin{equation}
A_{\mu}= \frac {1}{a^{2}e} \phi \wedge \partial_{\mu} \phi
+ \frac {1}{a} \phi A_{\mu},
\end{equation}
and the field strength can be written
\begin{equation}
{\cal F}_{a}^{\mu \nu}= \frac {1}{a} \phi_{a} {\cal F}^{\mu \nu}=
\frac {1}{a} \phi_{a}\, \frac {1}{a^{3}e} \phi \cdot (\partial^{\mu} \phi
\wedge \partial^{\nu} \phi) + \partial^{\mu}A^{\nu}- \partial^{\nu}A^{\mu}.
\label{eq:fstr}
\end{equation}

Besides, the equations of motion derived from the Georgi-Glashow
lagrangian (\ref{eq:LGG}), when combined
with condition (\ref{eq:Hvac}), reduce to Maxwell's equations
(\ref{eq:Mvac}), a fact
remembering us that outside the monopole the $SO(3)$ gauge theory can not be
distinguished from electromagnetic theory.

Lets us now calculate the magnetic flux through a closed surface $S$.
Using (\ref{eq:fstr}),
\begin{equation}
g=\int_{S} B \, dS = - \frac {1}{2ea^{3}} \int_{S}
\epsilon_{ijk} \phi \cdot (\partial^{j} \phi \wedge \partial^{k} \phi) d S^{i}.
\end{equation}
This expresion is invariant under smooth deformations of the Higgs field: for
a field variation \( \phi '= \phi + \delta \phi \), such that
\begin{equation}
\phi \cdot \delta \phi = 0,
\label{eq:ort}
\end{equation}
the variation term
\begin{eqnarray}
\delta [\phi \cdot (\partial^{j} \phi \wedge \partial^{k} \phi)] & = &
3 \delta \phi \cdot (\partial^{j} \phi \wedge \partial^{k} \phi) + \nonumber \\
& & \partial^{j}[\phi \cdot (\delta \phi \wedge \partial^{k} \phi)] -
\partial^{k}[\phi \cdot (\delta \phi \wedge \partial^{j} \phi )]
\end{eqnarray}
vanishes up since, on integration, the last two terms vanish by Stoke's
theorem,
and the first, as \(\partial^{j}\phi \wedge \partial^{k}\phi \) is parallel
to $\phi$ (because \(\partial^{i}\phi \bot \phi \)), is zero due to
(\ref{eq:ort}).
Therefore, small variations in $\phi$ do not modify the flux $g$. This
result can be extended to all changes which can be built from small
deformations;
these small deformations are called {\em homotopies\/}\footnote{The maps $\phi$
can
be divided into equivalence classes under homotopy, two maps being in the
same class if and only if they are continously deformable into each other
(homotopic).$\phi$ defines a non trivial element of the second homotopy
group $\Pi_{2}({\cal V})$:
\[\Pi_{2}(SU(2)/U(1)) \sim \Pi_{1}(U(1)) \sim Z . \]}.

If we write the magnetic flux as \( g=- 4 \pi N/e \), the integral $N$, defined
through
\begin{equation}
N= \frac {1}{4 \pi a^{3}} \int_{S} dS^{i} \frac {1}{2} \epsilon_{ijk}
\phi \cdot ( \partial^{j} \phi \wedge \partial^{k} \phi ),
\end{equation}
turns out to be the winding number mentioned above characterizing the map
$\phi$: the number of times $S^{2}$ is wrapped around $\cal V$ by $\phi$. It
must
therefore be an integer (a fact in agreement with Dirac quantization
condition).

\subsubsection{BPS States.}
As the monopole described so far is a smooth structure, a mass can be
calculated;
the expression for this mass was shown by Bogomolny to satisfy a simple bound:
\begin{equation}
M \geq a \mid g \mid.
\end{equation}
This bound can be saturated so, as was shown by Prasad and Sommerfield, we can
obtain a solution with
\begin{equation}
M = a \mid g \mid.
\end{equation}
Monopoles satisfying this bound are called Bogomolny-Prasad-Sommerfield (BPS)
mono\-poles \cite{BPS}.

The original ansatz by `t Hooft and Polyakov, leading to these simple
expresions,
was electrically neutral; however, the absence of electric charge is not a
necessary condition coming from spherical symmetry: it is possible to obtain
solutions, called {\em dyons\/}, containing both electric and magnetic charge.
Again, a simple bound holds for the mass of these BPS (saturated) states:
\begin{equation}
M=a(q^{2}+g^{2})^{1/2}.
\end{equation}

If we use the relation between electric and magnetic charge coming from the
assymptotic expression (1.29), $g=-4 \pi/e$, the mass formula for dyons with
$n_{e}$ units of electric charge, and $n_{m}$ units of magnetic charge, can
be written
\begin{equation}
M=\mid a\,e (n_{e}+ \tau_{0} n_{m}) \mid,
\label{eq:mas}
\end{equation}
where we have introduced a parameter $\tau_{0}$ containing the coupling:
\begin{equation}
\tau_{0} \equiv i \frac {4 \pi}{e^{2}}.
\end{equation}

\vspace{5 mm}

An important modification of equation (\ref{eq:mas}), coming from the CP
non invariant $\theta$-term in the lagrangian, was discovered by Witten
\cite{t}.
In non abelian gauge theories, pure gauge vacuum configurations define, if
we use the temporal gauge $A_{0}=0$, maps from the compactified space three
sphere $S^{3}$ into the gauge group $G$ (maps classified into
homotopy classes by $\Pi_{3}(G)$, which for $G=SU(N)$ is the set of integer
numbers, $Z$).

Instantons are euclidean field configurations that tunnel between vacua in
different topological classes. Denoting, in the temporal gauge $A_{0}=0$, by
$\mid n \! >$ the pure gauge vacuum corresponding to a pure gauge configuration
characterized by the value $n$ in $\Pi_{3}(G)$, the net effect of instantons
\cite{cl}
is to define the $\theta$-vacuum as the coherent state
\( \mid \theta \! >= \sum e^{i n \theta} \mid n \!>\),
with the tunneling amplitude \( < \! n \mid n+1 \!> \) given, in semiclassical
aproximation, by $e^{-S_{inst}}=e^{-8 \pi^{2}/g^{2}}$ ($S_{inst}$ is the
classical
action for the instanton). The generating function $<\! \theta \mid \theta \!>$
is then given by
\begin{equation}
< \theta \! \mid \theta \!> = \int dA \: exp[-\int {\cal L} + \frac {\theta}{32
\pi^{2}}
{\cal F}\tilde{{\cal F}}],
\label{eq:tt}
\end{equation}
with $\tilde{{\cal F}}$ the dual field tensor. Now, we should take into account
the effect of the $\theta$-parameter on the classical mass formula for the
magnetic monopole. Witten's result \cite{t} is that the monopole in the
presence
of
a non vanishing value of $\theta$ becomes effectively a dyon with electric
charge $\frac {\theta}{2 \pi}n_{m}$. This effect modifies the BPS mass formula
(\ref{eq:mas}) to
\begin{equation}
M= \mid ae \; (n_{e}+ \tau n_{m}) \mid,
\label{eq:mass}
\end{equation}
where now $\tau$ is defined by
\begin{equation}
\tau \equiv \frac {\theta}{2 \pi} + i \frac {4 \pi}{e^{2}}.
\label{eq:tx}
\end{equation}

The appearance in such a natural way of the ``complexified'' coupling constant
$\tau$ in the BPS mass formula is already a hint that some supersymmetry is
hiddenly governing monopole dynamics. In fact, as supersymmetry practitioners
know, the holomorphicity properties underlying non renormalization theorems are
intimately connected with the complexification (\ref{eq:tx}) of
the coupling constant \cite{smi} (more on this can be found on next section).

One further comment is the manifest symmetry of (\ref{eq:tt}) under the change
$\theta \rightarrow \theta + 2 \pi $. Witten's dyon effect provides a new
physical
flavour to this innocent symmetry: the transformation
$\theta \rightarrow \theta + 2 \pi$ changes in a non trivial way the induced
electric charge of the monopole:
\begin{equation}
\frac {\theta n_{m}}{2 \pi} \rightarrow \frac {\theta n_{m}}{2 \pi} + n_{m}
\label{eq:weff}
\end{equation}
This transformation will marry, as we will describe at the end of this section,
the old duality introduce by Dirac, to define the full duality group.

\subsubsection{Quantum Mass Formulas.}
The hidden supersymmetry that was trying to show up through the natural
appearance of the complexified coupling constant $\tau$ in the classical BPS
mass formula becomes manifest after the seminal comment by Olive and Witten
\cite{qm}, that points out that the BPS mass formula can be derived
directly from the $N\!=\!2$ supersymmetry algebra once we take into account the
existence of non vanishing central extensions in the Higgs phase. In fact, the
usual supersymmetry algebra
\begin{equation}
\{Q_{\alpha i},\bar{Q}_{\beta j}\}=\delta_{ij} \gamma^{\mu}_{\alpha
\beta}P_{\mu},
\> i,j=1,2,
\end{equation}
should be modified to include central terms \cite{HLS}:
\begin{equation}
\{Q_{\alpha i},\bar{Q}_{\beta j}\}=\delta_{ij} \gamma^{\mu}_{\alpha
\beta}P_{\mu} +
\delta_{\alpha \beta} U_{ij} + (\gamma _{5})_{\alpha \beta} V_{ij}, \> i,j=1,2.
\label{eq:Stop}
\end{equation}
The central terms in the above expression verify $U_{ij}=-U_{ji}$,
$V_{ij}=-V_{ji}$.

For the $N\!=\!2$ supersymmetric extension of the Georgi-Glashow model,
\begin{eqnarray}
{\cal L} & = & - \frac {1}{4}{\cal F}_{a}^{\mu \nu}{\cal F}_{a \mu \nu} +
\frac {1}{2} \bar{\psi}_{ai}i {\not{\!{\cal D}}} \psi_{a}^{i} +
\frac {1}{2} {\cal D}^{\mu}A_{a} {\cal D}_{\mu}A_{a} +
\frac {1}{2} {\cal D}^{\mu}B_{a}{\cal D}_{\mu}B_{a} + \nonumber \\
& & \frac {1}{2}e^{2} Tr [A,B][A,B] + \frac {1}{2} i \epsilon_{ij}
Tr ([\bar{\psi}^{i},\psi^{j}]A+[\bar{\psi}^{i},\gamma_{5}\psi^{j}]B),
\end{eqnarray}
and for non zero vacuum expectation value $<\!A\!>$, the central extensions
become respectively
\begin{equation}
U=<\!A\!>e \: \: \: \: \: \: \: \: \: \: V=<\!A\!>g,
\end{equation}
where g is the magnetic charge. Now, the algebra (\ref{eq:Stop}) can be seen to
imply, for the mass of each particle, the
relation
\begin{equation}
M^{2} \geq U^{2}+V^{2},
\label{eq:bbq}
\end{equation}
that is, the Bogomolny bound. With the notation used so far,
\begin{equation}
M^{2}\geq a^{2} (q^{2}+g^{2}).
\label{eq:Bb}
\end{equation}

The main interest of the previous result is that now we can claim that the
bound (\ref{eq:bbq}) has not only classical, but also quantum mechanical
meaning.
In fact, if supersymmetry is not dynamically broken, then we can be sure that
a formula like (\ref{eq:bbq}) will be exact, even after including all quantum,
perturbative and non perturbative, corrections. The only thing we need is to
use
the $N\!=\!2$ supersymmetric algebra of the effective theory obtained after
taking into account all quantum corrections.

$N\!=\!2$ supersymmetry does not only explain from a fundamental point of view
the
Bogomolny bound, but also clarifies the meaning of BPS saturated states: the
question
on when the above bound is saturated. Following the reasoning by Olive and
Witten, concerns the representations of the $N\!=\!2$ supersymmetry algebra.
An irreducible representation has $2^{N}$
helicity states for zero mass, and $2^{2N}$ states for nonzero mass, so we
might wonder whether the irreducible representations of the extended
supersymmetry
algebra should have four or sixteen states; it turns out that representations
with four helicity states (which are denoted ``small irreps'') are the ones
saturating the Bogomolny bound (\ref{eq:Bb}),
while it is not saturated for representations with sixteen states.

Particles getting mass by the Higgs mechanism ($W^{\pm}$, monopoles, dyons),
must be irreducible representations of the extended supersymmetry
algebra; in fact, for $<\!A\!> \neq 0$ the central terms are non vanishing.
Moreover, if they get mass by the Higgs mechanism, which does not change the
number of degrees of freedom, they must have $4$ helicity states, as massless
particles have, and therefore they will transform under $N\!=\!2$ supersymmetry
as small irreps, so that they will be BPS states.

\subsection{Duality.}
\label{sec:duality}
In this section we will use the symmetries of mass formula for BPS saturated
states to define the duality group. To do so, let us introduce some
notation to rewrite the mass formula (\ref{eq:mass}),
\[ M=\mid a\,e(n_{e}+ \tau n_{m}) \mid, \> \> \tau=i \frac {4 \pi}{e^{2}}
+ \frac {\theta}{2 \pi},\]
in a more convenient form: if we define
\begin{equation}
a\equiv a \cdot e, \> \> a_{D}\equiv \tau a,
\end{equation}
then the mass spectrum will be given by
\begin{equation}
M=\mid an_{e} + a_{D}n_{m} \mid.
\end{equation}
Now, we know that shifting the $\theta$ angle by $2 \pi$, \(\theta \rightarrow
\theta + 2 \pi \), should have no effect. Such a shift amounts to
\begin{equation}
\tau \rightarrow \tau +1
\label{eq:shift}
\end{equation}
or, in terms of our new variables,
\begin{equation}
a \rightarrow a \> \> \> a_{D} \rightarrow a + a_{D}.
\label{eq:s2}
\end{equation}
In order to make sure that the mass spectrum is not modified, (\ref{eq:s2})
should
be accompanied by the change
\begin{equation}
(n_{e},n_{m}) \rightarrow (n_{e}-n_{m},n_{m}).
\label{eq:nw}
\end{equation}

The shift (\ref{eq:shift}) is a fractional linear transformation,
\begin{equation}
\tau \rightarrow \frac {a \tau + b}{c \tau + d},
\label{eq:tfrac}
\end{equation}
with $ad-bc\neq 0$, where now $a=b=d=1$, $c=0$. In matrix form,
\begin{equation}
        \left( \begin{array}{c} a_{D} \\ a \end{array} \right) \rightarrow
        \left( \begin{array}{cc} 1 & 1 \\ 0 & 1 \end{array} \right)
        \left( \begin{array}{c} a_{D} \\ a \end{array} \right) =
        \left( \begin{array}{c} a + a_{D} \\ a \end{array} \right).
\end{equation}
The matrix appearing in the above expression is known as $T$:
\begin{equation}
        T \equiv \left( \begin{array}{cc} 1 & 1 \\ 0 & 1 \end{array} \right).
\end{equation}
It is an element of $Sl(2,Z)$, the group (``special linear group'') of
$2\times2$
matrices of unit determinant, with integer entries: \(T \in Sl(2,Z)\). This
special group is also known as the {\em (full) modular group\/}.

Notice from (\ref{eq:nw}) that the effect of $T$ on the mass spectrum is simply
Witten's effect (\ref{eq:weff}).

Now let us consider Dirac's electric-magnetic transformation (\ref{eq:Mdual}):
\begin{equation}
n_{e} \rightarrow n_{m} \> \> n_{m} \rightarrow - n_{e}.
\end{equation}
No change will appear in the mass formula if we also perform
\begin{equation}
a \rightarrow a_{D} \> \> a_{D} \rightarrow -a.
\end{equation}
Again, this is a fractional linear transformation, with $a=d=0$, $b=-1$, $c=1$:
\begin{equation}
\tau \rightarrow - \frac {1}{\tau}.
\end{equation}
In matrix form,
\begin{equation}
        \left( \begin{array}{c} a_{D} \\ a \end{array} \right) \rightarrow
        \left( \begin{array}{cc} 0 & -1 \\ 1 & 0 \end{array} \right)
        \left( \begin{array}{c} a_{D} \\ a \end{array} \right) =
        \left( \begin{array}{c} -a \\ a_{D} \end{array} \right),
\end{equation}
with
\begin{equation}
        S \equiv \left( \begin{array}{cc} 0 & -1 \\ 1 & 0 \end{array} \right)
        \in Sl(2,Z).
\end{equation}
The electric and magnetic variables exchange generated by $S$ leads to a
strong-weak coupling transformation,
\begin{equation}
S: \tau \rightarrow - \frac {1}{\tau}
\end{equation}
(Notice that it is \(\tau \equiv i \frac {4 \pi}{e^{2}} + \frac {\theta}{2 \pi}
\),
and not $e$, which is properly inverted.)

\vspace{5 mm}

Now we can combine Dirac's duality $S$ with Witten's effect as defined by $T$.
These two transformations generate the modular group $Sl(2,Z) \equiv \Gamma$
\cite{al},
an ubiquitous symmetry group in physics. This group is defined by the relations
\begin{equation}
S^{4}=\II \: \: \: \: \: \: \: \: \: \: (ST)^{3}=\II.
\end{equation}
We will call this group the {\em duality group\/}. The complexified coupling
constant $\tau$ is a number living in the complex upper half plane, once we
impose on it the physical constraint of positivity. The action of the duality
group on this plane is defined by (\ref{eq:tfrac}).

The existence of a fundamental domain for $Sl(2,Z)$ shows how difficult will
be to find a theory invariant under duality transformations, if such a theory
has non vanishing $\beta$-functions (i. e., non trivial running coupling
constant).
The necessary condition will be that the renormalization group trajectories be
concentrated in the fundamental region.

The Montonen-Olive conjecture, that the symmetry under the duality group
is an exact symmetry of some quantum theory, implies that strong coupling is
equivalent to the weak
coupling limit, with particles and solitons exchanged.

\vspace{5 mm}
When the $\beta$-function is zero, the
duality conjecture proposed by Montonen and Olive can be directly expressed
as the modular invariance of the partition function. Let us then
think of a theory with a lagrangian ${\cal L}(A,\psi,\bar{\psi},\phi;
e,\theta)$;
the Montonen-Olive duality conjecture states that the partition
function
\begin{equation}
{\cal Z}(e,\theta)=\int DA\,D\psi\,D\bar{\psi}\,D\phi\,
e^{-\int {\cal L}(A,\psi,\bar{\psi},\phi;e,\theta)} \equiv {\cal Z}(\tau)
\end{equation}
remains invariant under the action of the duality group; therefore, the
transformations generated by $T$ and $S$ must leave $\cal Z$ invariant:
\begin{eqnarray}
{\cal Z}(\tau) & = {\cal Z}(\tau + 1) \nonumber \\
{\cal Z}(\tau) & = {\cal Z}(-1/\tau)
\end{eqnarray}
Thus, invariance under the duality group
$Sl(2,Z)$ means that the partition function $\cal Z$ must be a modular form.

\begin{quote}A two dimensional
example of duality in a theory is that given by string $T$-duality, where
\[{\cal Z}(G,B)=\int DX exp [\int (G^{ij} \partial X^{i} \partial X^{j} +
\epsilon B^{ij} \partial X^{i} \partial X^{j} ) d^{2}z].\]
As the $\beta$-functions vanish, \(\beta_{G}=\beta_{B}=0\), \({\cal Z}(\tau)=
{\cal Z}(-1/\tau)\), with \(\tau \equiv i G + B\).

Candidates to four dimensional dual theories are those with vanishing
$\beta$-function:
$N\!=\!4$ supersymmetric Yang-Mills (the original place where Montonen-Olive
duality used to live), and $N\!=\!2$ supersymmetric $SU(N_{c})$ QCD with
$N_{f}\!=\!2N_{c}$.

In next section we will study the extension of duality given by Seiberg and
Witten
to the context of $N\!=\!2$ supersymmetric theories.
\end{quote}


\section{Duality and Effective Field Theories.}

\subsection{Flat Potentials and Classical Moduli.}

 The existence of flat potentials is generic in $N\!=\!1$ and $N\!=\!2$
supersymmetric
gauge theories, as well as in the low energy limit of string theories.  The
best
way to
characterize the special features of a gauge theory possessing a flat potential
is using
the concept of a {\em classical moduli \/}. In the previous section we have
introduced
the vacuum manifold $\cal V$, defined as the set of all the vacuum field
configurations.
Different points in $\cal V$ can be parametrized by the corresponding vacuum
expectation
values of the scalar fields entering into the theory. Given a generic point $ P
\in \cal V$
 we define the gauge group $H_{P}$ as the part of the gauge symmetry group $G$
of
the lagrangian consisting of symmetries of the vacuum state parametrized by
$P$. Different points $P$, $P'$ in $\cal V$ will describe the same physics if
we can reach $P'$ from $P$ by acting with some element of the gauge symmetry
group $G$, i.e., $P' \in \  G/H_{P}$ or, in other words, $P$ and $P'$ are
related by some Goldstone boson.

Now, we define the {\em moduli space} $\cal M$($\cal V$) as the
space of equivalence classes of vacua, where two vacua will be in the same
equivalence
class if they describe the same physics, i.e., if they are related by the
action
of Goldstone bosons. If the theory we start with possesses a flat potential
then
the moduli $\cal M$($\cal V$) will be a connected manifold with dimension
bigger
or equal to one.

A more geometrical description of the moduli can be done as follows. Given a
generic point $P \epsilon \cal V$ we can decompose the tangent space $T_{P}$
as:
\begin{equation}
T_{P}=T_{P}^{\cal G}\otimes T_{P}^{\cal M},
\label{eff1}
\end {equation}
where the generators of $T_{P}^{\cal G}$ are the Goldstone bosons, and the
generators of $T_{P}^{\cal M}$ are properly speaking the tangents to the moduli
 directions. The dimension of $T_{P}^{\cal G}$ is given by the dimension of the
homogeneous space $G/H_{P}$.

{\em Singularities \/} in $\cal M$($\cal V$) will appear at points $P$ where
jumps in the dimension of $T_{P}^{\cal G}$ take place. The meaning of these
singularities
is clear from the physical point of view, namely they correspond to points in
$\cal V$ where the symmetry of the vacuum $H_{P}$ changes.

It is clear, from the definition of $\cal M$($\cal V$), that in order to define
good coordinates we should use gauge invariant quantities. To fix ideas, let us
consider a concrete example\footnote{We follow the conventions of \cite{SW},
so the Higgs field is normalized so that its kinetic term is multiplied by
$1/4\pi \alpha = 1/g^{2}$}:
\begin{quote}
For $N\!=\!2$ supersymmetric Yang-Mills with gauge group $G\!=\!SU(2)$ the
potential
for the complex scalar field $\phi$ is given by
\begin{equation}
V(\phi)= \frac {1}{g^{2}} Tr [\phi , \phi^{\dag}]^{2},
\end{equation}
with $g$ the coupling constant. The flat direction is defined by
\begin{equation}
\phi= \frac {1}{2}a \sigma^{3},
\label{eq:flat}
\end{equation}
where $a$ is a complex parameter, and $ \sigma ^{3}$ is the diagonal Pauli
matrix.
For $a \neq 0$ the gauge symmetry is spontaneously broken to $U(1)$. Vacuum
states
corresponding to values $a$ and $-a$ are equivalent, since they are related by
the
action of the Weyl subgroup of $SU(2)$. A gauge invariant  parametrization of
$\cal M$($\cal V$) in this case can be defined in terms of the expectation
value
of the Casimir $Tr \phi ^{2}$:
\begin{equation}
u \equiv Tr \phi ^{2} = \frac {1}{2}a^{2}.
\label{eq:u}
\end{equation}
For $u\!=\!0$ we find a singularity of $\cal M$($\cal V$). In fact, at this
point there
is an enhancement of the gauge symmetry, and $H(u\!=\!0)\!=\!SU(2)$, while for
all
the other points, with $u \neq 0$, we have $H(u \neq 0)\!=\!U(1)$. In this
example,
the moduli $\cal M$($\cal V$) is simply the complex plane punctured at the
origin\footnote{The concept of moduli in the sense described above should be
familiar to practitioners of conformal field theory (CFT). Given a CFT its
moduli is generated by the set of truly marginal operators. The singularities
of the moduli correspond to those points where some relevant operator becomes
truly
marginal. The metric of the moduli space is known in CFT as the Zamolodchikov's
metric \cite{Zm}. In the context of string theory the flat directions of the
four
dimensional low energy physics are associated with the moduli of the CFT used
to characterize the internal space-time.}.
\end{quote}

\subsection{The Quantum Moduli.}

Generically we expect that the flat directions of the classical potential
disappear
once we take into account perturbative and non-perturbative quantum
corrections.
Notice that different points in the moduli $\cal M$($\cal V$) correspond
to inequivalent physical theories; for instance, for a theory with
spontaneously
broken gauge symmetry and non trivial moduli space $\cal M$($\cal V$), the
mass of the gauge vector boson will be different for different points in
$\cal M$($\cal V$). From physical grounds we expect that after including the
quantum corrections the vacuum degeneracy will be lifted and one particular
theory
will be selected. If this is not the case, and after including all quantum
corrections
the flat direction remains flat, then we will be able to define the moduli
$\cal M$$_{q}$($\cal V$) of the complete quantum theory; this manifold will
be called the {\em quantum moduli}.

The first question we should consider in such a case is in what way the
geometry of $\cal M$$_{q}$($\cal V$ ) will differ from that of the classical
moduli
$\cal M$($\cal V$). Clearly, the differences between $\cal M$($\cal V$) and
$\cal M$$_{q}$($\cal V$) will reside in the location and number of
singularities.
In fact, as we have already explained, singularities are associated with jumps
in the symmetry invariance of the vacuum; more concretely, they appear whenever
a charged particle in the spectrum becomes massless. The simplest example was
the
singularity of $\cal M$($\cal V$), for $SU(2)$ $N\!=\!2$ SYM, at the origin,
where
the gauge vector bosons $W^{\pm}$ become massless and the gauge invariance is
classically
restored.

When quantum mechanical effects are turned on, a classical singularity can
disappear if the associated massless particle is not quantum mechanically
stable. In the same way, some new singularities can appear whenever a classical
massive particle becomes massless once quantum corrections are taken into
account.

\vspace{5 mm}

But before entering into a detailed study of the singularities of the quantum
moduli
$\cal M$$_{q}$($\cal V$), we should consider the previous question of the very
existence of the quantum moduli \cite{Sth}:

\vspace{3 mm}

{\bf Theorem}\hspace{8 mm}{\em For $N\!=\!2$ supersymmetric Yang-Mills the
classical flat direction (\ref{eq:flat})
remains flat after quantum corrections, perturbative and non perturbative.}

\vspace{3 mm}

The proof of this statement goes as follows. First, we observe that the only
way
to generate a superpotential for the $N\!=\!1$ chiral matter superfield $\Phi $
is by breaking the extended $N\!=\!2$ to $N\!=1\!$. In fact, the most general
$N\!=\!2$ invariant theory is described by
\begin{equation}
{\cal L} _{eff}=Im \:[\: \int d^{2} \theta  \: d^{2} \bar{\theta} \frac {1}{2}
K(\Phi ^{a},
        \bar{\Phi}^{a}, V^{a}) + \int d^{2} \theta f_{ab}(\Phi)\, W^{a} \,
W^{b}
+ c \: c\:],
\label{eq:Leff}
\end{equation}
where $f_{a\,b}(\Phi)$ is a holomorphic function, and $K$ is a K\"{a}hler
potential. Quantum corrections will determine the specific form of $K$ and $f$
in (\ref{eq:Leff}). The second step in the proof uses the fact that Witten's
index $tr \:(-1)^{F}$,
for $N\!=\!2$ supersymmetric Yang-Mills, is different from zero \cite{index},
which automatically
implies that supersymmetry is not broken dynamically. Combining these two
facts,
the theorem follows.

\subsection{Effective Field Theory description of the Quantum \mbox{Moduli.}}
\label{sec:EFT}
Let us consider an $N\!=\!2$ supersymmetric gauge theory\footnote{Here we
consider
the case of a gauge group $SU(N)$.} possessing a non trivial quantum moduli
$\cal M$$_{q}$($\cal V$). For each point $P \in$$\cal M$$_{q}$($\cal V$), let
us denote by $S_{pec}(P)$ the corresponding mass spectrum. Some particles in
$S_{pec}(P)$ will become massive by the Higgs mechanism with respect to the
vacuum
expectation values which parametrize the point P in the flat direction. To
define an effective field theory at $P$ requires:
\begin{itemize}
        \item To split $S_{pec}(P)$ into light and heavy particles.
        \item To integrate the heavy particles.
\end{itemize}

Using the theorem discussed in the previous paragraph we can conclude that the
effective theory at $P$ will be described, up to higher derivative terms, by
a Lagrangian of the type (\ref{eq:Leff}) for a set of \mbox{$N\!=\!2$}
hypermultiplets $\Psi ^{i}(P)$ that describe the
light particles. More precisely, the K\"{a}hler potential $K$ and the
holomorphic function $f$ in (\ref{eq:Leff}) will be determined by the
integration of the
heavy modes, and the $N\!=\!2$ hypermultiplets entering into the lagrangian
will correspond to the light particles. Notice that we should split the
spectrum
into light and heavy modes in a way consistent with $N\!=\!2$ supersymmetry.

Now, we define effective field theory coordinates of the point $P \in$$\cal
M$$_{q}$($\cal V$)
by the expectation values of the scalar components of the $N\!=\!2$
hypermultiplets describing the light modes. Again, as a concrete example, let
us
consider the case of $SU(2)$ $N\!=\!2$ supersymmetric Yang-Mills: For a point
with a value of $u \neq 0$ the spectrum of massive particles is determined
by the Higgs mechanism, and the light modes will be described by one
$N\!=\!2$ hypermultiplet containing the $U(1)$ photon.

The question we should adress now is the range of validity of some
\mbox{`` EFT-coordinates ''}. Two are the general criteria we must take into
account:
\begin{enumerate}
        \item The range of validity of the split of the spectrum,
                \begin{equation}
                S_{pec}(P)=S_{pec}^{light}(P)\: \oplus \: S_{pec}^{heavy}(P)
                \end{equation}
        into light and heavy modes. We should require that when moving
        continously in $\cal M$$_{q}$($\cal V$) light particles go into
        light, and heavy into heavy. The effective field theory description
        around a point $P$ will break down whenever we reach a point in
        $\cal M$$_{q}$($\cal V$) such that a heavy particle becomes light.
        This is a similar phenomena to a spectral flow (see Figure~1).


        \begin{figure}[htbp]
        \centering

        \begin{picture}(400,170)

                \put(150,100){\line(1,0){140}} \put(150,97){\line(1,0){140}}
                \put(150,94){\line(1,0){140}} \put(150,91){\line(1,0){140}}

                \put(150,30){\line(1,0){140}} \put(150,33){\line(1,0){140}}
                \put(150,36){\line(1,0){140}} \put(150,27){\line(1,0){140}}

                \multiput(150,63)(4,0){35}{\line(1,0){2}}
                \multiput(220,20)(0,4){20}{\line(0,1){2}}
                \put(150,86.6){\line(3,-1){140}}

                \put(310,5){$\cal M$$_{q}$($\cal V$)} \put(80,92){heavy}
                \put(80,31){light} \put(100,110){$S_{pec}(P)$}
                \put(235,140){\vector(-1,-2){15}} \put(244,138){range of
validity}
        \thicklines
                \put(130,20){\line(1,0){220}} \put(150,20){\vector(0,1){100}}
                \put(290,20){\line(0,1){100}} \put(220,15){\line(0,1){10}}
                \multiput(220,15)(0,2){5}{\line(1,0){3}}

        \end{picture}
                \caption{Spectral flow in the quantum moduli.}
        \end{figure}
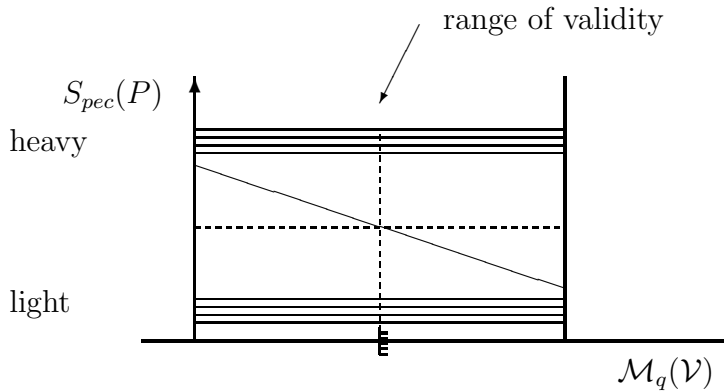


        The net effect, at the level of the effective field theory,
        of some heavy particle becoming light is,
        when this particle is charged with respect to some gauge group $G$,
that
        the corresponding effective coupling constant g develops a logarithmic
singularity,
        of the type
                \begin{equation}
                \frac {1}{g^{2}} \: \sim \: \ln (m),
                \label{eq:ln}
                \end{equation}
        with $m$ the mass of the light particle.
        \item The second general criteria for the validity of the effective
field
        theory description is, of course, to reduce ourselves to regions in
        $\cal M$$_{q}$($\cal V$) for which the corresponding effective field
theory
        is weakly coupled.
\end{enumerate}

The two criteria just described give us already some hints on what is going to
be the ``EFT-geometry'' of the quantum moduli space $\cal M$$_{q}$($\cal V$).
In
fact, we are going to need different EFT-coordinates to cover the whole
quantum moduli. In the overlaping regions, where one should pass from some
EFT-coordinates to others, we will need to use a change of EFT-coordinates
satisfying some conditions: namely, to be isometries of the quantum moduli
Zamolodchikov's metric. It is at this point where the concept of duality will
play an important role, as we will see in the rest of this notes.

Before ending this introductory section on the EFT-geometry of the quantum
moduli,
let us briefly come back to the criterion $1$, in connection with the structure
of
singularities of $\cal M$$_{q}$($\cal V$). Recall that singularities in
$\cal M$$_{q}$($\cal V$) were also associated with some massive particle in the
spectrum becoming massless. From our previous discussion, it becomes clear that
the structure of singularities of $\cal M$$_{q}$($\cal V$) is telling us how
many different local EFT-coordinates we will need to use in order to cover the
whole quantum moduli.

\subsection{Duality and EFT-Geometry.}
\label{sec:DeffG}

\subsubsection{Wilsonian Effective Theories.}

It has become traditional, after the work of the russian school \cite{smi}, to
differentiate
between the effective action interpreted as the $1PI$ generating functional,
and what is known as the Wilsonian effective action. The Wilsonian effective
action, $S_{W}(\mu)$, is defined by integrating the vacuum loops with virtual
momentum $p> \mu$. Thus, the difference between the wilsonian effective action
and the $1PI$ generating functional depends on the infrared region, where $p<
\mu$.
Denoting by $g_{W}(\mu)$ the wilsonian effective coupling, and using for $\mu$
the mass scale determined by the Higgs mechanism, i.e., the vacuum expectation
value
of the scalar field, we can define
\( 1/{g_{W}^{2}(\mu)}\)
as a function on $\cal M$$_{q}$($\cal V$). In most cases it is important to
differentiate between the wilsonian effective coupling $g_{W}(\mu)$ and the
effective
coupling $g_{eff}(\mu)$, defined as the coefficient of the corresponding
$1PI$ vertex for external momentum equal to $\mu$. In particular, the physical
$\beta$-function, $\beta(g)$, is defined for $g_{eff}(\mu)$, and not for
$g_{W}(\mu)$.

To see the difference between $g_{eff}(\mu)$ and $g_{W}(\mu)$ it is convenient
to use the so called Konishi anomaly \cite{Kan}. To illustrate the phenomena,
let us just
consider the simpler case of $N\!=\!1$ SQED. The wilsonian effective action
is
\begin{equation}
{\cal L}_{W}= \frac {1}{4\, g_{W}^{2}(\mu)} \, \int d^{2} \theta \, W\, W \: +
\:
\frac {Z(\mu)}{4} \, \int d^{4} \theta \, ( \bar{T} e^{V} T \, + \, \bar{U}
e^{-V} U),
\label{eq:LW}
\end{equation}
with
\begin{equation}
\frac {8 \pi ^{2}}{g_{W}^{2}(\mu)} = \frac { 8 \pi ^{2}}{g_{0}^{2}} +
2 \ln \left( \frac {\Lambda}{\mu} \right),
\end{equation}
where $\Lambda$ is the ultraviolet cut-off. The D-term in (\ref{eq:LW}) can be
written,
using standard superfield notation, as follows:
\begin{equation}
\int d^{4} \theta ( \bar{T} e^{V} T + \bar{U} e^{-V} U ) =
        - \frac {1}{2} \int d^{2} \theta \bar{D}^{2} ( \bar{T} e^{V} T +
        \bar{U} e^{-V} U ).
\end{equation}
By using the Konishi anomaly relation,
\begin{equation}
Z \bar{D}^{2} ( \bar{T} e^{V} T + \bar{U} e^{-V} U )=
\frac {1}{2 \pi ^{2}} W^{2},
\end{equation}
we easily get the extra ``infrared'' contribution to $g_{eff}(\mu)$:
\begin{equation}
\frac {8 \pi ^{2}}{g_{eff}^{2}(\mu)} =
\frac { 8 \pi ^{2}}{g_{0}^{2}} + 2 \ln \left( \frac {\Lambda}{\mu} \right) -
2 \ln Z(\mu) =
\frac {8 \pi^{2}}{g_{W}^{2}(\mu)} - 2 \ln Z(\mu)
\label{eq:geffW}
\end{equation}
and, in this way, the explicit relation between the wilsonian effective
coupling
and the effective coupling for $N\!=\!1$ SQED.

\subsubsection{The $N\!=\!2$ Prepotential.}

One of the main characteristics of $N\!=\!1$ supersymmetric theories, which is
at the origin of non renormalization theorems, is the holomorphic dependence
of the wilsonian effective coupling on the scale $\mu$. As it can be observed
from the previous argument based on the Konishi anomaly, this is not in general
true for the effective coupling constant $g_{eff}(\mu)$ , which contains
pieces of an infrared origin coming from $D$-terms in the lagrangian which are
not holomorphic. This phenomenon is the
field theory analog of the string holomorphic anomaly.

For $N\!=\!2$ supersymmetric theories holomorphic constraints are stronger than
for $N\!=\!1$. In fact, a generic $N\!=\!2$ effective theory can be completely
described, in $N\!=\!2$ superfield notation, by means of the lagrangian
\begin{equation}
{\cal L} = Im \int d^{4} \theta {\cal F}_{eff} ( \Psi)
\label{eq:Lpre}
\end{equation}
with $\Psi$ representing the set of $N\!=\!2$ hypermultiplets, and
$\cal F$$_{eff}$($\Psi$) the prepotential, which is a holomorphic function.
To pass from (\ref{eq:Lpre}) to (\ref{eq:Leff}), we use
\begin{equation}
f_{ab}(\Phi) = \frac {\partial ^{2}{\cal F}}{\partial \Phi ^{a} \partial \Phi
^{b}},
\label{eq:cp}
\end{equation}
\begin{equation}
K( \Phi^{a},\bar{\Phi}^{b}, V) = \partial _{a}
{\cal F}\cdot (e^{V})_{ab} \bar{\Phi}^{b}.
\end{equation}

Notice that the whole $N\!=\!2$ lagrangian (\ref{eq:Lpre}) is, in $N\!=\!2$
superfield
notation, of the same type as the $N\!=\!1$ $F$-term superpotential, and
therefore we can extend the $N\!=\!1$ non renormalization theorems of the
superpotential to the whole $N\!=\!2$ lagrangian.

\subsubsection{Zamolodchikov's metric of the Quantum Moduli.}

Let us now come back to the quantum moduli \MqV\ .As we have discussed in
Section~\ref{sec:EFT}, we parametrize points in \MqV\  by the expectation
value of the scalar component of the hypermultiplet $\Psi(P)$ used to
describe the light modes in $S_{pec}(P)$. Moreover, the effective lagrangian
is given by $\cal F$$_{eff}$($\Psi (P)$), where the effective prepotential
is obtained integrating the massive (heavy) modes in $S_{pec}(P)$.

Now, we can use a $\sigma$-model inspired approach to the geometry of \MqV\ .
Based on relation (\ref{eq:cp}), and taking now into account that we are using
the
scalar components of the chiral multiplets $\Phi$ as the coordinates of \MqV\ ,
we can use the holomorphic function $f_{ab}(\Phi)$ to define the
Zamolodchikov's metric of \MqV\ . The physical meaning of this metric is now
specially clear from equation (\ref{eq:Leff}). Namely, and for the simplest
case
of a one complex dimensional quantum moduli, we get:
\begin{equation}
\frac { 4 \pi}{g_{W}^{2}(\mu)}=Im \: f(\Phi) \; \; \; \; \; \; \; \; \; \; \;
\;
\frac {\theta_{W}(\mu)}{2 \pi}=Re \: f(\Phi),
\label{eq:Wtau}
\end{equation}
where, as usual, $\mu$ here refers to the spectation value of the scalar part
in the multiplet $\Phi$, and we have introduced an effective ``wilsonian''
$\theta$-parameter as the real part of f($\Phi$).

{}From (\ref{eq:Wtau}), the relation between the Zamolodchikov's metric on
\MqV\
and the
renormalization group of the theory is manifest.

\subsubsection{Dual Coordinates.}

We will now come back to the notation used in section~1, besides working, for
simplicity, with a one complex dimensional quantum moduli. Denoting then by
$a$ the expectation value of the scalar field, the tree level prepotential for
the scalar component is given by
\begin{equation}
{\cal F}^{(0)}(a) = \frac {1}{2} \tau ^{(0)}a^{2},
\end{equation}
with \( \tau^{(0)} =  \frac {i 4 \pi}{g_{(0)}^{2}} + \frac {\theta_{(0)}}{2
\pi}\). The
tree level value $\theta_{(0)}$ is equal to zero. Now, using the
same notation as in section~\ref{sec:duality}, we introduce the
 {\em dual variable} $a_{D}$ as:
\begin{equation}
a_{D} \equiv \tau^{(0)} a = \frac {\partial {\cal F}^{(0)}}{\partial a}.
\label{eq:amag}
\end{equation}
The generalization of (\ref{eq:amag}) for the effective field theory defined by
${\cal F}_{eff}(a)$ is just
\begin{equation}
a_{D} \equiv \frac {\partial {\cal F}_{eff}(a)}{\partial a}.
\label{eq:Amgen}
\end{equation}

A physical way to check if definition (\ref{eq:Amgen}) of the dual variable is
meaningful
would be computing the mass of BPS states of the effective theory or, in
other words, to find the central extensions of the $N\!=\!2$ supersymmetric
algebra for the effective field theory defined by $\cal F$$_{eff}(a)$. In
fact, the mass formula for the BPS-saturated states of the effective theory
is given by
\begin{equation}
M^{2}(n_{e},n_{m}) = \: \mid a \, n_{e} \, + \,
\frac {\partial {\cal F}_{eff}}{\partial a} \, n_{m} \mid ^{2},
\end{equation}
in agreement with definition (\ref{eq:Amgen}).

Summarizing, at each point $P \in $\MqV\ we have introduced the following set
of
geometrical objects:
\begin{itemize}
        \item[{\it i)}] The effective field theory coordinate $a(P)$, defined
        by the expectation value of the scalar component of the hypermultiplet
        $\Psi(P)$, which describes the light modes in $S_{pec}(P)$.
        \item[{\it ii)}] The effective prepotential $\cal F$$_{eff}$$(a(P))$
obtained
        by integrating the heavy modes in $S_{pec}(P)$.
        \item[{\it iii)}] The {\em dual coordinate}
        \( a_{D}= \frac {\partial {\cal F}_{eff}(a(P))}{\partial a}\), in terms
        of which we reproduce the BPS mass formula derived from the centrally
extended
        supersymmetric algebra of the effective theory.
        \item[{\it iv)}] The Zamolodchikov's metric
        \(g(P)= \frac {\partial ^{2} {\cal F}_{eff}(a(P))}{\partial  a^{2}}\)
        of \MqV\  at the point $P$.
\end{itemize}

\subsubsection{Duality Transformations.}
\label{sec:DT}
In terms of the dual coordinate $a_{D}(P)$, the Zamolodchikov's metric can
be written as follows:
\begin{equation}
d s^{2}= Im \: \frac {\partial ^{2} {\cal F}_{eff}}{\partial a \, \partial
\bar{a}}
\: da \, d \bar{a} = Im \: d \bar{a} \,d a_{D},
\end{equation}
which is manifestly invariant under the $S$-duality transformation
\begin{equation}
S: \left( \begin{array}{c}
        a_{D}  \\ a \end{array} \right) \longrightarrow \left(
\begin{array}{cc}
                                                                 0 & 1 \\
                                                                -1 & 0
                                                               \end{array}
\right) \left( \begin{array}{c} a_{D} \\ a \end{array} \right) =
        \left( \begin{array}{c}
                a \\ -a_{D}
               \end{array} \right).
\label{eq:Sd}
\end{equation}
Moreover, in the dual variables the metric $g(P)$ becomes
\begin{equation}
g^{D}(P) = - \frac {1}{g(P)}.
\label{eq:Dmetric}
\end{equation}

The physical meaning of the dual coordinates $a_{D}(P)$ can be easily
understood in the general effective field theory framework we have used in our
previous discussion. Let us denote by $S_{pec}^{(l)}(P)$ the light part of
the spectrum that we are describing by means of the $N\!=\!2$ hypermultiplet
$\Psi(P)$, with scalar component $a(P)$. Now, we can formally define
$S_{pec}^{(l\mid D)}(P)$ as the dual of $S_{pec}^{(l)}(P)$. Particles in
$S_{pec}^{(l\mid D)}(P)$ are related to the ones in $S_{pec}^{(l)}(P)$ by
interchanging electric with magnetic charge. Therefore, in the case of
$N\!=\!2$ $SU(2)$ SYM, $S_{pec}^{(l)}(P)$ is described by the $N\!=\!2$
hypermultiplet of the unbroken $U(1)$ photon, and $S_{pec}^{(l \mid D)}(P)$
will
be described by a new hypermultiplet $\Psi^{D}(P)$, containing a ``dual''
photon.
The dual coordinate $a_{D}(P)$ will represent the scalar component of the
$N\!=\!2$
hypermultiplet $\Psi^{D}(P)$.

Equation (\ref{eq:Dmetric}) implies that if the effective field theory for
$\Psi(P)$ is
weakly (strongly) coupled, then the effective field theory for $\Psi^{D}(P)$
will be strongly (weakly) coupled. Using now that the duality transformation
(\ref{eq:Sd})
is an isometry of the Zamolodchikov's metric, we can try to use duality to
extend, beyond the weak coupling regime, the range of validity of a set of
EFT-coordinates.

\subsection{$N\!=\!2$ Non Renormalization Theorems.}
The popular way to prove non renormalization theorems in supersymmetric
theories is using the {\em multiplet of anomalies} argument \cite{man}. This
argument is based on the Adler-Bardeen theorem for the $U(1)$ axial anomaly,
and
the fact that by supersymmetry the axial and the conformal currents are
in the same hypermultiplet. From these two facts, formally follows that the
$\beta$-function is saturated by one loop contributions. This argument is
known to be wrong for $N\!=\!1$ supersymmetric theories, the reason being
that the supersymmetric partner of the conformal anomaly is not the one to
which the Adler-Bardeen theorem applies. The physical origin of this problem
is the same already discussed in section~\ref{sec:DeffG}, concerning the
differences
between the wilsonian and the effective coupling constant. In fact, a
``wilsonian $\beta$-function'' would be saturated by one loop corrections;
however, this is not the case for the standard $\beta$-function, as can be
easily observed from equation (\ref{eq:geffW}). For $N\!=\!2$ supersymmetric
theories,
the situation changes dramatically, and the multiplet of anomalies argument
produces the right result \cite{ire}. Starting from equation (\ref{eq:Lpre})
for
the effective
lagrangian, the form of $\cal F$$_{eff}(\Psi)$ can be fixed using:
\begin{itemize}
        \item[{\it i)}] Holomorphy of the effective prepotential, and
        \item[{\it ii)}] The $U(1)$ axial anomaly.
\end{itemize}

{}From the tree level prepotential \( {\cal F}^{(0)}(\Psi)=
\frac {1}{2} \tau^{(0)} \Psi ^{2} \), we fix the $U(1)$ $R$-charges of the
$N\!=\!2$ hypermultiplet:
\begin{equation}
R(\Psi) = 2.
\end{equation}
The $U(1)$ axial anomaly implies that $\cal L$$_{eff}$ is transforming
under the $U(1)$ axial transformations as
\begin{equation}
\delta_{\alpha}{\cal L}_{eff} =
\frac {\alpha}{4 \pi ^{2}}{\cal F} \tilde{{\cal F}}.
\label{eq:Lvar}
\end{equation}
Using (\ref{eq:Lpre}), (\ref{eq:Lvar}),
and the condition of holomorphy we derive the non renormalization theorem:
\begin{equation}
{\cal F}_{eff}(\Psi) = \frac {1}{8 g^{2}} \Psi ^{2}\;
[1+ \frac {g^{2}}{4 \pi^{2}} \ln \left( \frac {\Psi^{2}}{\Lambda^{2}} \right)],
\label{eq:Seff}
\end{equation}
which implies for the effective coupling defined by
\begin{equation}
Im \: \frac {\partial ^{2} {\cal F}_{eff}}{\partial \Phi \, \partial \Phi}
\end{equation}
the following renormalization group relation:
\begin{equation}
\frac {1}{g_{eff}^{2}(a)} = \frac {1}{g_{0}^{2}} +
\frac {1}{4 \pi ^{2}} \ln \left( \frac {a^{2}}{\Psi^{2}} \right),
\label{eq:grel}
\end{equation}
where we have already denoted by $a$ the scalar part of the $N\!=\!2$
superfield $\Psi$.

For $N\!=\!1$ supersymmetric theories, the previous way to prove the non
renormalization theorems can be directly applied to the superpotential
$F$-term of the lagrangian, which is also constrained by holomorphicity and
$R$-symmetries. The difference with $N\!=\!2$, concerning the coupling
constant, and therefore the non renormalization theorems for the
$\beta$-function, is that in $N\!=\!1$ theories the effective coupling
constant gets contributions through the Konishi anomaly from $D$-terms
in the lagrangian, which are not constrained by holomorphicity.

\subsection{The Singularity at Infinity.}
\label{sec:Sinf}

For $N\!=\!2$ supersymmetric $SU(2)$ Yang-Mills we can use
$\cal F$$_{eff}(\Psi)$, as given by equation (\ref{eq:Seff}), to define the
effective
field theory for the light modes, i.e., for $\Psi$ representing the $N\!=\!2$
gauge field of the photon. From equation (\ref{eq:grel}), we know that this
effective
field theory will be ``reasonable'' in the neighbourhood of $a=\infty$,
where the theory is weakly coupled. We can now compactify the moduli space
\MqV\  by adding the point at infinity; therefore, with this compactification,
the point at $\infty$ appears as a singularity in \MqV\ .In fact, this
singularity has a quantum origin, as it comes from the logarithmic term in
equation (\ref{eq:grel}).

To understand the physical meaning of this singularity, we can use the dual
coordinate introduced in equation (\ref{eq:Amgen}). From (\ref{eq:Seff})
and (\ref{eq:Amgen}), we get
\begin{equation}
a_{D}= \frac {2 i a}{\pi} \ln \left( \frac {a}{\Lambda} \right) +
\frac {i a}{\pi},
\label{eq:aD}
\end{equation}
which implies that at the point at $\infty$ the mass of the magnetic monopole
becomes infinite.

The effective field theory described by (\ref{eq:Seff}) represents the weak
coupling
regime of the electric light modes, once we have integrated the particles
becoming massive by the Higgs mechanism. The parametrization of $a$ in terms
of the gauge invariant coordinate $u$, can be defined by equation (\ref{eq:u}):
\begin{equation}
a(u)=\sqrt{2 \, u}.
\label{eq:uinv}
\end{equation}
Using (\ref{eq:aD}) and (\ref{eq:uinv}),
we can compute the monodromy induced by the singularity
at $\infty$, by looping in the physical parameter $u$:
\begin{eqnarray}
    a(e^{2 \pi i}u) & = & - a(u) \nonumber \\
a_{D}(e^{2 \pi i}u) & = & - \frac { 2 i a}{\pi} \ln \left( \frac {a}{\Lambda}
\right) -
\frac {i a}{\pi} + 2a = -a_{D} + 2a.
\end{eqnarray}
With the matrix notation also used in section~1, we get:
\begin{equation}
\left( \begin{array}{c} a_{D} \\ a \end{array} \right) \longrightarrow
        \left( \begin{array}{cc}
        -1 & 2 \\ 0 & -1 \end{array} \right)
        \left( \begin{array}{c} a_{D} \\ a \end{array} \right) =
        M_{\infty} \left( \begin{array}{c} a_{D} \\ a \end{array} \right),
\end{equation}
which in terms of the $Sl(2,Z)$ generators is given by:
\begin{equation}
M_{\infty}= P T^{-2},
\label{eq:Pap}
\end{equation}
where $P=- \II$.

{}From equation (\ref{eq:Seff}), and the definition of the Zamolodchikov's
metric,
we get
the transformation of $g(P)$ under the monodromy $M_{\infty}$:
\begin{equation}
g(e^{2 \pi i} P) = \frac { -g \, + \, 2}{-1}
\end{equation}
which from (\ref{eq:Wtau}) simply means
\begin{equation}
\theta_{eff} \rightarrow \theta_{eff} - 4 \pi,
\end{equation}
a perfectly nice symmetry of the effective theory. This result is in fact
general:
{\em any monodromy around a singularity of \MqV\  defines an exact symmetry of
the
quantum theory\/}.

\begin{quote}
Before going into a more detailed study of the physical meaning of this
symmetries, it would be adecuate to introduce the following classification of
the different elements in $Sl(2,Z)$:
\begin{description}
        \item[Classical symmetries] Diagonal matrices in $Sl(2,Z)$.
        \item[Perturbative symmetries] Matrices \( \left( \begin{array}{cc}
                                                                a & b \\
                                                                c & d
                                                           \end{array} \right)
\),
        with $c=0$.
        \item[Non perturbative symmetries] Matrices \( \left( \begin{array}{cc}
                                                                a & b \\
                                                                c & d
                                                                \end{array}
\right) \),
        with $c\neq0$.
\end{description}
Perturbative symmetries correspond to shifts in $\theta$
\( ( \theta \rightarrow \theta + 2 \pi n ) \), while {\em non perturbative
symmetries connect strong with weak coupling }.
\end{quote}

\subsection{Strong Coupling Regime: heuristic approach.}

For $u$ close to $\Lambda$, the dinamically generated scale of the theory,
we reach the strong coupling regime, so around
this point in \MqV\  we should look for a new set of weakly coupled light
modes.
{}From the classical formula for the `t Hooft--Polyakov monopole, we can expect
that magnetic monopoles are good candidates for defining the relevant light
modes,
weakly coupled, in terms of which the effective field theory description of
the strong coupling region in \MqV\  can be defined.

Let us asume that the mass of the monopole vanishes at $u=\Lambda$. In this
case, we know that the dual coupling constant defining the coupling of the
magnetic monopole to the dual photon will become singular at the point
$\Lambda$
(see equation (\ref{eq:ln})). The singularity will go as the logarithm of the
monopole mass, which from the BPS mass formula we know is given by the value
of $a_{D}$\footnote{Notice that the contribution of the monopole to the one
loop vacuum polarization of the dual theory should be computed with an
ultraviolet cutoff $\Lambda$ of the order $m_{W}$, since the size of the
monopole is $1/m_{W}$.}. Based on this simple argument, we can guess the form
of
the effective
prepotential $\cal F$$_{eff}^{D}(a_{D})$, describing the region of \MqV\
around
$\Lambda$ in terms of the dual coordinate $a_{D}$; namely,
\begin{equation}
{\cal F}_{eff}^{D}(a_{D}) \sim a_{D}^{2} \ln a_{D},
\label{eq:presing}
\end{equation}
with \(a_{D}(u) \mid _{u=\Lambda} = 0\). In the neighbourhood of $\Lambda$
we write \(a_{D}(u) \sim ( u - \Lambda) \). From (\ref{eq:presing}) we can now
define, using
equation (\ref{eq:Amgen}), the dual coordinate $a$ by:
\begin{equation}
a= \frac {\partial {\cal F}_{eff}^{D}}{\partial \, a_{D}} = a_{D}(u) \ln
a_{D}(u)
+a_{D}(u).
\label{eq:asing}
\end{equation}
Notice that the behaviour of $a$, given by (\ref{eq:asing}), is very different
to the
behaviour (\ref{eq:uinv}) defined by the classical Higgs mechanism in the weak
coupling
regime, already indicating that a different phase is governing the strong
coupling region.

\subsection{Montonen-Olive Duality in $N\!=\!2$ Theories.}
\label{sec:l}

{}From the previous discussion we can get some insight on the structure of
monodromy matrices associated with singularities of the quantum moduli
space. For BPS stable particles with charges $(0,n_{e})$, $(n_{m},0)$,
the two ingredients we shall use to find the corresponding monodromy
matrix at the point where the particle becomes massless are: i) The logarithmic
singularity in the effective coupling constant associated to the charge of
the particle (see equation (\ref{eq:ln})), and ii) The BPS mass formula.
In terms of variables $a(u)$, $a_{D}(u)$ we find for an ``electric''
singularity, i. e., a $(0,1)$ massless particle, the following behaviour:
\begin{equation}
a_{D}(u) \sim a(u) \ln a(u) + a(u),
\label{eq:aDn}
\end{equation}
with $a(u) \rightarrow 0$ at the singular point. In the same way, for a
``magnetic'' singularity associated with a massless $(1,0)$ particle we
will get:
\begin{equation}
a(u) \sim a_{D}(u) \ln a_{D}(u) + a_{D}(u),
\label{eq:an}
\end{equation}
with $a_{D} \rightarrow 0 $ at the singular point. In a graphical
representation (\ref{eq:aDn}) and (\ref{eq:an}) correspond to one loop
Feynman diagrams in which the particle running inside the loop is the one that
becomes massless at the singular point and the (two) external legs represent
the $U(1)$ photon, in the case of (\ref{eq:aDn}), or the dual $U(1)$ photon
for (\ref{eq:an}).

Originally, Montonen-Olive dualty was propposed as a symmetry with respect
(in Georgi-Glashow model) to the interchange of $W^{\pm}$ charged vector bosons
and `t Hooft-Polyakov magnetic monopoles. At tree level, this conjecture can
be described in terms of the Feynman diagram identity depicted in Figure~2.


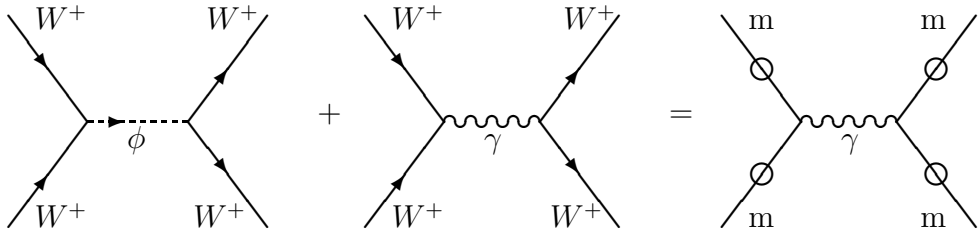
\begin{figure}[htbp]
\centering

\begin{picture}(400,170)
\thicklines
        \put(20,80){\line(3,-4){15}} \put(5,100){\vector(3,-4){15}}
        \put(20,40){\line(3,4){15}}  \put(5,20){\vector(3,4){15}}

        \multiput(35,60)(4,0){10}{\line(1,0){2}}
        \put(45,60){\vector(1,0){4}}

        \put(73,60){\vector(3,-4){15}} \put(73,60){\vector(3,4){15}}
        \put(88,80){\line(3,4){15}}    \put(88,40){\line(3,-4){15}}

        \put(15,95){$W^{+}$} \put(15,20){$W^{+}$} \put(80,95){$W^{+}$}
        \put(75,20){$W^{+}$} \put(50,50){$\phi$}  \put(122,60){+}

        \put(140,100){\vector(3,-4){15}} \put(170,60){\line(-3,4){15}}
        \put(170,60){\line(-3,-4){15}}   \put(140,20){\vector(3,4){15}}

        \put(172,60){\oval(4,4)[t]}
        \put(176,60){\oval(4,4)[b]}
        \put(180,60){\oval(4,4)[t]}
        \put(184,60){\oval(4,4)[b]}
        \put(188,60){\oval(4,4)[t]}
        \put(192,60){\oval(4,4)[b]}
        \put(196,60){\oval(4,4)[t]}
        \put(200,60){\oval(4,4)[b]}
        \put(204,60){\oval(4,4)[t]}

        \put(206,60){\vector(3,4){15}} \put(206,60){\vector(3,-4){15}}
        \put(221,80){\line(3,4){15}}   \put(221,40){\line(3,-4){15}}

        \put(150,95){$W^{+}$} \put(150,20){$W^{+}$} \put(185,50){$\gamma$}
        \put(215,95){$W^{+}$} \put(210,20){$W^{+}$} \put(255,60){=}

        \put(275,20){\line(3,4){15}} \put(275,100){\line(3,-4){15}}
        \put(305,60){\line(-3,-4){15}} \put(305,60){\line(-3,4){15}}

        \put(307,60){\oval(4,4)[t]}
        \put(311,60){\oval(4,4)[b]}
        \put(315,60){\oval(4,4)[t]}
        \put(319,60){\oval(4,4)[b]}
        \put(323,60){\oval(4,4)[t]}
        \put(327,60){\oval(4,4)[b]}
        \put(331,60){\oval(4,4)[t]}
        \put(335,60){\oval(4,4)[b]}
        \put(339,60){\oval(4,4)[t]}

        \put(341,60){\line(3,4){30}} \put(341,60){\line(3,-4){30}}

        \put(290,80){\circle{8}} \put(290,40){\circle{8}}
        \put(356,80){\circle{8}} \put(356,40){\circle{8}}

        \put(285,95){m} \put(285,20){m} \put(350,95){m} \put(350,20){m}
        \put(320,50){$\gamma$}
\end{picture}

       \caption{Feynman diagram for monopoles.}
\end{figure}


It is well known that this conjecture is not true for $N\!=\!0$ theories. In
the
case of the $N\!=\!2$ supersymmetric extension of the Georgi-Glashow model the
main problem this conjecture faces is that the $W^{\pm}$ vector bosons and
the magnetic monopoles transform under different representations of $N\!=\!2$
supersymmetry (vector bosons live in a vector multiplet, with spin one,
while the magnetic monopole is an hypermultiplet with $s=1/2$). Independently
of this general argument we can try to check the Montonen-Olive conjecture
already at the level of the monodromy matrices, just comparing the one
associated with massless $W^{\pm}$ and that generated by massless magnetic
monopoles. Denoting this monodromy matrices respectively by $M_{W}$ and
$M_{m}$, it is easy to convince ourselves, just using (\ref{eq:aDn}) and
(\ref{eq:an}), that both monodromies are certainly different. Moreover,
$M_{W}$ will be generated by the $T$ generator of $Sl(2,Z)$, while $M_{m}$
will contain the $S$ generator. At this point, the best we can do to generalize
Montonen-Olive duality conjecture to $N\!=\!2$ supersymmetric gauge theories
with a one dimensional quantum moduli will be to look for some extra monodromy
matrix $M$ such that:
\begin{equation}
M_{W}=M_{m} M.
\label{eq:Mex}
\end{equation}
As we will see in next section, this is in fact what happens in $SU(2)$ super
Yang-Mills with $M$ the monodromy matrix corresponding to the singularity
generated by a massless dyon. In fact, the general result for $N\!=\!2$
supersymmetric gauge theories with a one dimensional quantum moduli is:
\begin{equation}
M_{W}=M_{m} \prod_{i} M_{R_{i}(m)},
\label{eq:Mb}
\end{equation}
where $M_{R_{i}(m)}$ are the monodromy matrices corresponding to singularities
associated with massless $R_{i}(m)$ particles, where by $R_{i}(m)$ we denote
BPS stable states obtained from the monopole by acting with the set of unbroken
global $R$-symmetries of the theory.

Notice that the only way to satisfy the Montonen-Olive duality conjecture,
interpreted as the identity $M_{W}=M_{m}$, is when either the massless
$W^{\pm}$ or the massless monopole produce a singularity in the one loop
diagrams, i. e., when the theory has a vanishing $\beta$-function.

Equations (\ref{eq:Mex}) and (\ref{eq:Mb}) will characterize the way
Montonen-Olive duality is extended to $N\!=\!2$ supersymmetric theories with
non vanishing $\beta$-function. In the next lecture, we will work out the
previous picture in more detail.


\section{Exact Results and Coupling to Gravity.}

\subsection{Singularities and Phases.}

In this section we will reduce ourselves to the study of $N\!=\!2$
$SU(2)$ super Yang-Mills. As we have seen in the previous section, this theory
possess a flat potential which is not lifted by quantum corrections, and
therefore
a one dimensional quantum moduli \MqV\ . A gauge invariant parametrization
of \MqV\ is defined by means of the Casimir coordinate $u=Tr \; \phi ^{2}$.
In the Higgs phase the electrically charged vector bosons $W^{\pm}$ have
a mass given by $a(u)\,n_{e}$ ($n_{e}=1$), with $a(u)$ the vacuum expectation
value of the Higgs field, i. e., $\sqrt{2u}$. If we maintain ourselves in
the Higgs phase, the geometry of \MqV\ can be described by the effective
field theory prepotential ${\cal F}_{eff}(a(u))$, where we integrate all
the massive (heavy) particles, with the mass of these particles obtained
by the standard Higgs mechanism. Singularities in the Higgs phase will
appear whenever one of these massive particles becomes massless. This
singularity will introduce some logarithmic dependence of the dual variable
$a_{D} \equiv \frac {{\cal F}_{eff}(a(u))}{\partial a(u)} $ on $a(u)$.
The origin of this logarithmic singularity is, of course, the coupling of the
$a(u)$ field to a Higgsed massive state that becomes massless at the
singularity.

The monodromy matrices in the Higgs phase, $\{ M_{i}^{Higgs} \}$, coming from
\begin{equation}
        \left( \begin{array}{c} a_{D}(u) \\ a(u) \end{array} \right)
\rightarrow
        M_{i}^{Higgs} \left( \begin{array}{c} a_{D}(u) \\ a(u) \end{array}
\right) =
        \left( \begin{array}{c} a_{D}(u e^{2 \pi i}) \\ a(ue^{2 \pi i})=-a
                        \end{array} \right),
\end{equation}
will be of the type
\begin{equation}
        M_{i}^{Higgs}= \left( \begin{array}{cc} a & b \\ 0 & -1 \end{array}
        \right) \in Sl(2,Z)
        \label{eq:MHiggs}
\end{equation}
for some integer values of $a$ and $b$ depending on the quantum corrections
to the effective prepotential.

{}From (\ref{eq:MHiggs}) we immediately derive the following general result:
\begin{itemize}
        \item[{\it $R_{1}$}] On the Higgs phase of \MqV\ , for $N\!=\!2$
        $SU(2)$ super Yang-Mills, the monodromy group generated by the
singularities
        is abelian, i. e., it is generated by $T$ and $P$.
\end{itemize}
The previous result in particular means (see section~\ref{sec:duality}) that
the exact quantum symmetry defined by the monodromy group reduces to simply
the well known symmetry \( \theta \rightarrow \theta + 2 \pi n, \: n \in Z \).

The question we should address now is whether the whole quantum moduli \MqV\
is in the Higgs phase.

\subsubsection{Holomorphicity and Abelian Monodromy.}
\label{sec:Abmon}

A simple holomorphicity argument can prove to us that the quantum moduli
\MqV\ contains more than the Higgs phase. The argument goes as follows: First
of all we define the coupling constant, in the way described in the previous
Section, as
\begin{equation}
Im \; \tau(a(u)),
\label{eq:cceff}
\end{equation}
with
\begin{equation}
\tau(a(u))= \frac {\partial ^{2} {\cal F}_{eff}(a(u))}{\partial a^{2}}.
\end{equation}

Obviously, with respect to the abelian monodromy of type (\ref{eq:MHiggs}),
and on the Higgs phase defined by \( a(u) \sim \sqrt {2u} \), the effective
coupling constant (\ref{eq:cceff}) will be single valued or, in other words,
{\em globally defined\/}\footnote{This can be trivially derived from the
way the abelian subgroup of $Sl(2,Z)$ generated by $T$ is acting on $\tau$
(equation (\ref{eq:shift}):
\[ T: \tau \rightarrow \tau + 1, \]
which implies that the imaginary part of $\tau$ is unchanged.}.

{}From the general structure of $N\!=\!2$ supersymmetric theories we know that
the prepotential is an holomorphic function; therefore, \( Im \, \tau (a(u)) \)
is harmonic. Now, we only need to remember some basics in complex analysis,
namely the well known theorem that states that an harmonic function can only
reach the maximum at the boundary of its domain of definition. Therefore, if we
impose positivity of the effective coupling constant (\ref{eq:cceff}), then
the Higgs phase can only correspond to a local region of the quantum moduli
space. Summarizing, we have obtained the following second result:
\begin{itemize}
        \item[{\it $R_{2}$}] The quantum moduli of $N\!=\!2$ supersymmetric
        Yang-Mills can not be globally described in terms of the Higgs phase
        variables.
\end{itemize}

Geometrically, we are learning that $a(u)$ and $a_{D}(u)$ should be interpreted
as sections on \MqV\ . It should be already clear, from our discussion in
the previous sections on the way the duality group is defined, that these
sections are sections of a two dimensional vector bundle on \MqV\ with
structure
group, acting on the fibre, the duality group $Sl(2,Z)$.

The simplest (and naive) way to interpret $R_{2}$ would be noticing that at
the origin we are out of the Higgs phase, because at that point the gauge
group is unbroken; however, this comment, as we will see in a moment, is
wrong. In fact, at the origin the charged particles that become massless are
the gauge vector bosons. The monodromy at that point should be such that the
spectrum vector ($n_{e}=1, \; n_{m}=0$) is invariant under its action. This
already means that this monodromy should be part of the abelian subgroup
generated by $T$ and $P$, and therefore will not help us in solving the
problem of positivity of $Im \; \tau(a(u)) $. The reader should notice that
we are generically calling Higgs phase the whole region of \MqV\ where the
Higgs parametrization \( a(u) \sim \sqrt{2u} \) is correct.

\subsubsection{Duality and Phases.}
\label{sec:phases}

Now we would like to have an heuristic and simple minded way to
understand the previous phenomena, namely the existence of more than one phase
on \MqV\ . If we describe the Higgs phase by $a(u) \sim \sqrt{2u}$, we can
try to use the dual description to get some insight on what can be the
physical origin of the failure of this Higgs relation. To do that we can try
to work out, with the dual variable $a_{D}(u)$, an effective theory
${\cal F}_{eff}^{D}(a_{D}(u))$, and to define $a(u)$ by
$\frac {\partial {\cal F}_{eff}^{D}}{\partial a_{D}(u)}$. The variable
$a_{D}(u)$, as we have discussed in section~\ref{sec:DT}, describes what we
can call the ``dual'' photon; this photon is coupled to magnetically charged
particles which are represented under $N\!=\!2$ supersymmetry by $N\!=\!2$
matter hypermultiplets (the spin of the magnetic monopole is $1/2$). From
this picture we can now expect a logarithmic dependence in $a(u)$ (see
equation (\ref{eq:aD})), in contrast to the Higgs dependence
$a(u) \sim \sqrt{2u}$, whenever the dual photon is coupled to massless magnetic
hypermultiplets. Therefore, the other phase on which the theory is living can
correspond to having massless magnetic monopoles, something that, as we will
see, can be described as a dual (magnetic) Higgs phase.

\subsection{Seiberg-Witten Solution for $N\!=\!2$ $SU(2)$ Supersymmetric
Yang-Mills.}

Based on the previous discussion, we can introduce a set of rules that can be
used to derive the exact geometry of the quantum moduli space.
\begin{itemize}
        \item[{\it Rule 1}] Monodromy condition.
        We will assume that the quantum moduli is compactified by adding the
        point at $\infty$. In these conditions, we should impose
        \begin{equation}
        \prod M_{i} = \II,
        \label{eq:Mon}
        \end{equation}
        where the product in (\ref{eq:Mon}) is over the whole set of
singularities\footnote{Notice
        that monodromies are associated with non contractible paths in the $u$
        plane. The product in (\ref{eq:Mon}) is determined by combining paths
        with the same base point.}. If the moduli space has dimension
        bigger than one, we will assume that
        singularities always define codimension one regions.
\end{itemize}
\begin{itemize}
        \item[{\it Rule 2}] Positivity of the coupling constant.
        In the one dimensional case this condition, together with the
holomorphy
        of the prepotential, is enough to prove the existence of more than one
        Higgs phase.
\end{itemize}
\begin{itemize}
        \item[{\it Rule 3}] Global $R$-symmetries.
        Global $R$-symmetries are generically broken, by non perturbative
        instanton effects, to some discrete residual subgroup. We will require
        that singularities of the quantum moduli are mapped into singularities
        by the action of these global $R$-symmetries.
\end{itemize}
\begin{itemize}
        \item[{\it Rule 4}] BPS stability of massless particles.
        To each singular point $P_{i}$ we associate a massless charged particle
        characterized by the charge vector $(n_{m}^{(i)},n_{e}^{(i)})$. This
        vector should satisfy
        \begin{equation}
        (n_{m}^{(i)} \: \: n_{e}^{(i)}) \; M_{i} = (n_{m}^{(i)} \: \:
n_{e}^{(i)}),
        \label{eq:mdcond}
        \end{equation}
        and must correspond to a stable BPS particle.
\end{itemize}

We will now use the previous set of rules to build up the exact quantum moduli
for $N\!=\!2$ $SU(2)$ super Yang-Mills.
The starting point is of course the singularity at $\infty$ that we have
already described (see section~\ref{sec:Sinf}). The corresponding
monodromy is
\begin{equation}
M_{\infty}= \left( \begin{array}{cc} -1 & 2 \\ 0 & -1 \end{array} \right).
\end{equation}
By {\em Rule 2\/}, and the argument in~\ref{sec:Abmon} and~\ref{sec:phases},
we assume the existence of one point where the magnetic monopole becomes
massless; this point can be characterized by the dinamically generated scale
$\Lambda$. From (\ref{eq:aDn}) we derive
\begin{equation}
M_{\Lambda}=\left( \begin{array}{cc} 1 & 0 \\ -2 & 1 \end{array} \right).
\end{equation}
(From (\ref{eq:mdcond}) it can be now verified that the particle becoming
massless at
$u=\Lambda$ is in fact a magnetic monopole of charge one.)

Now we should take into account {\em Rule 3\/}, and to act with the residual
global $R$-symmetries on the point $u=\Lambda$. For $N\!=\!2$ $SU(2)$
super Yang-Mills, and when the vacuum expectation value is zero, the instanton
induces an effective vertex with eight external gluinos; when the vacuum
expectation value is different from zero, the number of zero modes reduces to
four, two coming from the fermionic partner of the gauge field, $\lambda$,
and two coming from the fermionic partner of the scalar field, $\psi$.

Therefore the $Z_{8}$ residual $R$-symmetry breaks to $Z_{4}$ for generic
points in \MqV\ or, in other words, $Z_{8}$ is spontaneously broken to
$Z_{4}$, and thus the orbit of the $R$-symmetry is $Z_{2}$. Hence, if we have
a singularity at $\Lambda$, we should have another singularity at the point
$-\Lambda$ (as $Z_{2}:u \rightarrow -u $).

This defines the minimal solution compatible with the previous set of rules,
where
$M_{-\Lambda}$ is obtained from condition (\ref{eq:Mon}):
\begin{equation}
M_{-\Lambda}=\left( \begin{array}{cc} -1 & 2 \\ -2 & 3 \end{array} \right).
\end{equation}
Using (\ref{eq:mdcond}) we now observe that the particle that becomes massless
at the point $-\Lambda$ is a dyon, which is again a BPS stable particle.

The $Z_{2}$ transformation can be implemented in the $u$ plane by a matrix $A$:
\begin{equation}
M_{- \Lambda}=A M_{\Lambda} A^{-1}.
\label{eq:Aconj}
\end{equation}
It is easy to observe that any $A$ of the form
\begin{equation}
A=T M_{\Lambda}^{r}
\end{equation}
provides a solution to (\ref{eq:Mon}). However, only for $r=1$
we get a solution satisfying $A^{2}=-\II$, which in particular implies that
after the action of $Z_{2}^{2}$ the stable monopole at $u=\Lambda$ will
become an antimonopole, a fact related to the existence of $P$ in
$M_{\infty}$ (see (\ref{eq:Pap})).

\subsection{Some Comments on Seibeg-Witten Solution.}

The most impressive implication of Seiberg-Witten solution is certainly that
the classical singularity at the origin is not there when quantum mechanical
effects are taken into account. This, in particular, means that over the
whole quantum moduli the gauge symmetry is $U(1)$, with no point where the
(full $SU(2)$) gauge symmetry is restored. This fact will be crucial to
connect Seiberg-Witten quantum moduli with type II strings.

A way to understand why the origin is not a singular point of the quantum
moduli
can be motivated by our discussion in section~\ref{sec:l} on Montonen-Olive
duality for $N\!=\!2$ supersymmetric theories. In fact, Seiberg-Witten
solution can be directly derived from equation (\ref{eq:Mex}). In other words,
we can start formally with a quantum moduli possesing only the singularities
at $\infty$ and at the origin. The singularity at the origin should be
generated
by massless gauge vector bosons, and therefore will be given by $M_{W}$.
Next, we impose the Montonen-Olive duality relation (\ref{eq:Mex}) for
$N\!=\!2$ theories and we obtain Seiberg-Witten solution or, equivalently,
the split of the classical singularity at the origin into the two
singularities at the points $\pm \Lambda$.

\vspace{5 mm}

A different way to approach the meaning of the two singularities at $\pm
\Lambda$
is in the framework of the `t Hooft, Polyakov and Mandelstam theory of
confinement. For asymptotically free theories the confinement phase is
expected to correspond to unbroken gauge symmetry (the classical singularity
at the origin) but with a vacuum characterized, as a ``dual'' BCS
superconductor,
by a non vanishing magnetic order parameter which, very likely, will
require massless magnetically charged objects (the two quantum singularities at
$\pm \Lambda $).

\subsubsection{The Abelian Confinement Argument.}

Let us now summarize the main steps of Seiberg-Witten confinement argument.
\begin{itemize}
        \item[{\it i)}] The quantum moduli of $N\!=\!1$ supersymmetric $SU(2)$
        Yang-Mills is a discrete set of two points, related by a $Z_{2}$ global
        $R$-symmetry transformation. This result comes from the exact
computation
        of Witten's index, $tr \:(-1)^{F}$, which in this particular case is
two
        (if the gauge group is taken to be $SU(N)$, the value of the index is
$N$).
        \item[{\it ii)}] The two vacua of $N\!=\!1$ supersymmetric $SU(2)$
        Yang-Mills are characterized by a non vanishing expectation value
        $<\! \lambda \lambda \!>$, a gaugino condensate \cite{bf}.
        \item[{\it iii)}] Instantons contribute to a non vanishing expectation
value
        $<\! \lambda \lambda (x) \: \lambda \lambda (y) \!>$ \cite{smifr}. As,
from supersymmetric
        Ward identities we know that this expectation value is independent of
        $\mid x-y \mid$, the gaugino condensate can be derived from the
instanton
        contribution to $<\! \lambda \lambda (x) \: \lambda \lambda (y) \!>$
        using cluster descomposition or, equivalently, assuming the existence
of
        a mass gap in the $N\!=\!1$ theory.
        \item[{\it iv)}] Adding a soft supersymmetry breaking term $m \: tr
\phi^{2}$
        to the $N\!=\!2$ theory, and using the decoupling theorem, we can
        define $N\!=\!1$ $SU(2)$ super Yang-Mills as the corresponding
effective
        low energy field theory. Once we add this soft breaking term, we
        lift the flat direction of the $N\!=\!2$ potential.
        \item[{\it v)}] The two vacua defining the quantum moduli of the low
        energy effective $N\!=\!1$ theory should correspond to two points of
the
        $N\!=\!2$ quantum moduli. Moreover, the existence of a mass gap in the
        $N\!=\!1$ theory implies that the massless $U(1)$ photon must become
        massive by some Higgs mechanism at these two points. The only candidate
        that can play the role of the Higgs field is the massless monopole
        at the points $\pm \Lambda$ that will higgs the dual ``magnetic''
photon.
        \item[{\it vi)}] This dual Higgs mechanism explains, in the dual analog
        of BCS superconductivity, the confinement and the mass gap of the
        $N\!=\!1$ theory.
\end{itemize}

More quantitatively, we can define a superpotential $W(M)$ for the monopole
field:
\begin{equation}
W(M)=m \; tr \phi^{2} + a_{D} M \tilde{M},
\end{equation}
with the term $a_{D}M \tilde{M}$ describing the coupling of the monopole to
the ``dual'' photon, as required by $N\!=\!2$ supersymmetry. On the quantum
moduli we can rewrite $W(M)$ using the fact that $a_{D}=a_{D}(u)$ as
\begin{equation}
W(M)=m \; u(a_{D}) + a_{D} M \tilde{M}.
\end{equation}
The vacuum defined by $dW=0$ is characterized, if \( \frac {\partial u}
{\partial a_{D}} \neq 0 \), by the magnetic order parameter
$<\! M \!> \neq 0$ which will induce, by the dual Higgs mechanism, the mass
gap of the $N\!=\!1$ theory. Notice that the confinement picture we are
presenting here takes only into account the abelian gauge symmetry.
The analog in $N\!=\!0$ quantum field theory is Polyakov's compact
quantum electrodynamics.

\subsection{Geometrical Interpretation.}

The monodromy group generated by $M_{\Lambda}$, $M_{-\Lambda}$ and $M_{\infty}$
is the group $\Gamma_{2}$ of unimodular matrices congruent to the identity
mod(2):
\begin{equation}
\Gamma_{2} \equiv \left\{ \begin{array}{c} \left( \begin{array}{cc} a & b \\ c
&
d \end{array}
                  \right) \in Sl(2,Z), a \equiv b \equiv 1 \: mod (2),
                  b \equiv c \equiv 0 \: mod (2) \end{array} \right\}.
\end{equation}
As explained above this group defines the exact quantum symmetry of the theory,
which in particular implies that we can reduce the upper half complex plane
$H$, parametrizing the coupling constant and the $\theta$-parameter, to the
$\Gamma_{2}$ fundamental domain $H/\Gamma_{2}$. This fundamental domain has
a nice interpretation in algebraic geometry that we will describe now in
very qualitative terms: Let us define an elliptic curve $E_{u}$ as the
vanishing
locus of a cubic polynomial in $\IP^{2}$,
\begin{equation}
W(x,y,z;u)=0,
\label{eq:Ecurve}
\end{equation}
and let us denote by $\tau(u)$ the corresponding elliptic modulus.
Singularities
of the curve defined by (\ref{eq:Ecurve}) will appear at values of $u$ for
which
\begin{equation}
W=\frac {\partial W}{\partial x}=\frac {\partial W}{\partial y}=\frac
{\partial W}{\partial z}=0.
\label{eq:Esing}
\end{equation}
Let us now denote by $\Gamma_{M}$ the monodromy group of the map $\tau(u)$ at
these singular points. By construction, $\tau(u)$ and $\Gamma_{M} \tau(u)$
should correspond to the same elliptic curve, and therefore
\( \Gamma_{M} \subset Sl(2,Z) \), the modular group of an elliptic (genus one)
curve. The quotient group
\begin{equation}
Sl(2,Z)/\Gamma_{M}
\label{eq:Q}
\end{equation}
will map among themselves the singular points solution to equation
(\ref{eq:Esing}).
In fact, all of them should correspond to the boundary of the moduli space
of complex structures of the defining elliptic curve. We can now characterize
the quotient (\ref{eq:Q}) as the set of transformations $x \rightarrow x'$,
$y \rightarrow y'$, $z \rightarrow z'$ such that
\begin{equation}
W(x',y',z';u)=f(u)W(x,y,z;u),
\label{eq:trans}
\end{equation}
i. e., as changes of local coordinates that can be compensated by a change of
the
moduli parmeter $u$.

\vspace{5 mm}

In order to reproduce Seiberg-Witten solution in the previous framework we
should find a cubic polynomial $W(x,y,z;u)$ \cite{itec} with solutions to
(\ref{eq:Ecurve})
at the points $u=\infty, \pm \Lambda$, and with monodromy group $\Gamma_{M}=
\Gamma_{2}$. It is easy to check that
\begin{equation}
W(x,y,z;u)=-zy^{2}+x(x^{2}- \Lambda^{4} z^{2})-uz(x^{2}- \Lambda^{4} z^{2})=0
\end{equation}
which defines the elliptic curve
\begin{equation}
E_{u}: \: \: y^{2}=(x- \Lambda^{2})(x+ \Lambda^{2})(x-u)
\label{eq:SWcurve}
\end{equation}
possesses singularities at precisely the points $u=\infty, \pm \Lambda$, and
that the group of transformations (\ref{eq:trans}) is isomorphic to
$Sl(2,Z)/\Gamma_{2}$, which indirectly means, by the argument above, that
the monodromy group of the corresponding map $\tau(u)$ is $\Gamma_{2}$.

For the curve (\ref{eq:SWcurve}) the map $\tau(u)$ can be easily defined as:
\begin{equation}
\tau(u)=\frac {\oint_{\gamma_{1}}\lambda_{1}}{\oint_{\gamma_{2}}\lambda_{1}}
\: \: \: \: \: \: \: \lambda_{1}=\frac {dx}{y} \in H^{1,0}(E_{u},C),
\label{eq:tauq}
\end{equation}
with $\gamma_{1}$, $\gamma_{2}$ an homology basis of $E_{u}$. From the
definition
of $\tau(u)$ in terms of the prepotential ${\cal F}(a(u))$ we get:
\begin{equation}
\tau(u)=\frac {da_{D}/du}{da/du}.
\label{eq:taupre}
\end{equation}
To match (\ref{eq:tauq}) and (\ref{eq:taupre}) we can try to look for a one
form $\lambda$ in $H^{1}(E_{u},C)=H^{0,1}\oplus H^{1,0}$ such that
\begin{equation}
a_{D}(u)=\oint_{\gamma_{1}}\lambda, \: \: \: \: \: \: \: \:
a(u)=\oint_{\gamma_{2}}\lambda,
\end{equation}
and
\begin{equation}
\frac {d \lambda}{d u} =f(u) \lambda_{1}
\end{equation}
for some function $f(u)$. To fix this function we can use the asymptotic
behaviour of $a_{D}(u)$ and $a(u)$ at the points $u=\pm \Lambda, \infty$,
which we have already derived in the previous sections (see equations
(\ref{eq:aD}) and (\ref{eq:uinv}) for the asymptotic behaviour at $\infty$,
and (\ref{eq:asing}) for the behaviour at the point $+ \Lambda$; recall that
the behaviour at $- \Lambda$ is derived from the one at $\Lambda$ by acting
with the residual global $Z_{2}$ $R$-symmetry). We leave as an exercise to
check that the correct one form $\lambda$ is given by
\begin{equation}
\lambda =\frac {\sqrt{2}(\lambda_{2}-u \lambda_{1})}{2 \pi},
\: \: \: \: \: \: \: \: \lambda_{2}=\frac {x dx}{y}.
\label{eq:SWsol}
\end{equation}

The natural connection between $N\!=\!2$ supersymmetry and algebraic
geometry should be understood in the framework of Picard-Fuchs equation.
{}From the $N\!=\!2$ supersymmetry transformations we can derive the following
set of relations \cite{sgeom}:
\begin{eqnarray}
d_{u}V & = & U, \nonumber \\
D_{u}U & = & C_{uuu}G_{u \bar{u}}^{-1}\bar{U}, \nonumber \\
d_{u}\bar{U} & = & 0,
\label{eq:rkr}
\end{eqnarray}
where we have introduced
\begin{eqnarray}
V & \equiv & (a,a_{D}), \nonumber \\
G_{u \bar{u}} & \equiv & Im \; \tau(u), \nonumber \\
C_{uuu} & \equiv & \frac {d \tau}{du} \left( \frac {da}{du} \right)^{2},
\nonumber \\
D_{u} & \equiv & d_{u}-\Gamma_{u}, \nonumber \\
\Gamma_{u} & \equiv & G_{u \bar{u}}^{-1}(d_{u}G_{u \bar{u}}),
\end{eqnarray}
with $\tau(u)$ and $a_{D}$ defined in the usual way in terms of the
prepotential
${\cal F}(a(u))$. The rigid K\"{a}hler relations (\ref{eq:rkr}) can be
organized
in the form of a differential equation; using relation (\ref{eq:taupre}) this
differential equation becomes the Picard-Fuchs equation for the periods of
the algebraic curve (\ref{eq:SWcurve}).

\vspace{5 mm}

Before finishing this section we would like to add some general comments.
\begin{itemize}
        \item[{\it C1}] Solution (\ref{eq:SWsol}) gives us the exact geometry
        of the quantum moduli space or, in other words, the exact prepotential.
        In physical terms this is equivalent to knowing the effective low
        energy lagrangian up to higher (bigger than two) derivative terms.
        \item[{\it C2}] The exact solution (\ref{eq:SWsol}) contains all the
        information concerning instanton effects.
\end{itemize}

\subsection{The Stringy Approach to the Quantum Moduli.}

Once you have the algebraic geometrical description of the quantum moduli space
it is difficult to resist (mostly if you have been exposed during the last
decade to the stringy way of thinking high-energy physics) the temptation
to interpret the results in stringy terms. In this spirit, it would be very
natural to put in parallel Seiberg-Witten solution for $N\!=\!2$
supersymmetric gauge theories with the effective field theory interpretation
of the special K\"{a}hler geometry of the moduli space of complex structures
of some Calabi-Yau manifold. The analogy is certainly more than formal. If
you consider a type II string compactified on some Calabi-Yau manifold $X$
the special K\"{a}hler geometry on the moduli ${\cal M}(X)$ of complex
structures can be interpreted in terms of an effective $N\!=\!2$ supergravity
with as many $U(1)$ gauge fields as the dimension of ${\cal M}(X)$, which
is given by the Betti number $b_{2,1}$ of the manifold $X$.

The first similarity with the quantum moduli of $N\!=\!2$ supersymmetric gauge
theories is the appearance of only $U(1)$ gauge fields, a consequence of
choosing a type II superstring. The formal analogy goes on in the sense that
$N\!=\!2$ local supersymmetry implies the Picard-Fuchs equation for the periods
of the top form on the Calabi-Yau manifold. Moreover, the exact quantum
duality symmetry of the effective supergravity theory describing the geometry
on ${\cal M}(X)$ is given by the $T$-duality, in string language, of the
Calabi-Yau manifold. The map $\tau(u)$ becomes, in this picture, the mirror
map mapping the moduli ${\cal M}(X)$ of complex structures into its mirror,
the moduli of K\"{a}hler structures, which we describe by $\tau$.
Singularities in ${\cal M}(X)$, which are known as conifold singularities,
should be the stringy parallel of the singularities of the quantum moduli.
The previous set of analogies can be temptatively summarized in the
following set of ``translation'' rules:

\begin{center}
\begin{tabular}{ccc}

        {\bf A: QFT Language}         &                   & {\bf B: String
Language} \\
                                      &                   &
    \\
        Quantum moduli of $N\!=\!2$   & $\leftrightarrow$ & Moduli of complex
structures \\
        supersymmetric gauge theory.  &                   & of some Calabi-Yau
manifold X \\
                                      &                   & with respect to
which we compactify \\
                                      &                   & a type II
superstring. \\
        Singularities (monopoles).    & $\leftrightarrow$ & Conifold
singularities. \\
        $\tau(u)$ map.                & $\leftrightarrow$ & Mirror map. \\
        $\Gamma_{M}$ monodromy group. & $\leftrightarrow$ & $T$-duality. \\
        $E_{u}$ curve.                & $\leftrightarrow$ & Calabi-Yau manifold
X.
\end{tabular}
\end{center}

\vspace{5 mm}

The previous set of analogies is very suggestive, but presents severe
difficulties
of interpretation. First of all we should notice that column B can be
interpreted
as describing the quantum moduli of some $N\!=\!2$ supergravity theory, and
therefore in order to pass to column A we need to work out some way to decouple
gravitational effects. In second place, the analogy between the quantum moduli
of $N\!=\!2$ supersymmetric gauge theories and the moduli, in column B, of
complex structures of some Calabi-Yau manifold, presents the problem that,
as we have discussed in the previous sections, the moduli of $N\!=\!2$
$SU(2)$ super Yang-Mills is not the moduli of complex structures of $E_{u}$,
but something bigger containing extra geometrical information on $E_{u}$, an
effect that is related to the way Montonen-Olive duality is realized in
$N\!=\!2$ supersymmetric theories with non vanishing $\beta$-function.
Another problem of our set of translation rules is to unravel the meaning
of the conifold singularity as associated with some massless charged
hypermultiplet\footnote{The
meaning of the conifold singularities constituted for a while a serious
puzzle. Recently, Strominger \cite{bh} has propposed to interpret these
singularities, in
perfect parallel with the approach of Seiberg and Witten, as coming from a BPS
stable massless charged hypermultiplet, that in string theory has the
interpretation
of a charged black hole. This conjecture can be checked once we consider
string loop corrections near the conifold point. Using the topological
twisted version we can compute the topological amplitudes $F_{g}$ at genus $g$
using the Kodaira-Spencer theory \cite{atft}. In this approach, $F_{g}$ is
given
by:
\[ F_{g}=(V_{mmm})^{2g-2}P^{3g-3}, \]
with $P$ the Kodaira-Spencer propagator, and $V_{MMM}$ the vertex for the
massive excitations. Using the relations between the structure constants
$C_{ijk}$, the propagator $P$ and the vertex,
\[ \partial_{l}C_{ijk} \sim P (V_{ttM})^{2}, \]
we easily get,
\[F_{g} \sim \frac {[\partial^{3}_{t} C_{ttt}]^{2g-2}}
{[\partial_{t}C_{ttt}]^{3g-3}}, \]
for $C_{ttt}$ the three point function corresponding to the marginal
directions defined by the $t$-direction of the moduli. Now, we observe from
the above relation that the tree level information of the special geometry of
the
moduli of complex structures, which determines $C_{ttt}$ and, therefore, the
string tree level conifold singularity, implies, for instance for $F_{1}$,
a logarithmic singularity that admits Strominger's interpretation of a
massless black hole running in the loop, in perfect analogy with our discussion
of the singularities of Seiberg-Witten quantum moduli.}.
And a final and most urgent problem is, of course, to figure out what can the
classical supergravity theory be whose quantum moduli is precisely described
by the moduli of complex structures of some Calabi-Yau manifold $X$.

Let us comment on this last issue. For a Calabi-Yau manifold $X$ with
$b_{2,1}=r$ we can naively think that the moduli ${\cal M}(X)$ is the
quantum moduli of a supergravity theory with gauge group $G$ spontaneously
broken to $U(1)^{r}$, with $r$ the rank of $G$. However, this is not the
correct answer because of the special role played by the dilaton field in
string theory. If, for instance, we consider that ${\cal M}(X)$ is the
quantum moduli of some classical moduli, described by ${\cal M}(Y)$, of
some different manifold $Y$, we need to count with a way to control the
quantum corrections on the theory compactified on the manifold $Y$ that
produces the quantum moduli ${\cal M}(X)$. This will be impossible if we
consider a type II compactification on $Y$ because in type II strings the
dilaton, which is counting the string loop effects, is in a hypermultiplet
and therefore does not interfere with the dynamics of vector multiplets,
which are the ones describing the geometry of the moduli space of
complex structures. The only possible way out is of course compactifying
on $Y$ a heterotic string. In this case the dilaton appears as a vector
hypermultiplet and therefore we have some chances that quantum corrections
of the heterotic string compactified on $Y$ sum up into the quantum
moduli space ${\cal M}(X)$. In this framework the second Betti number of $X$
should be equal to $r+1$, the rank of the gauge group of the supergravity
theory,
$G$, plus one extra vector field, corresponding to the dilaton of the heterotic
string. Pairs of Calabi-Yau manifolds $(X,Y)$ such that ${\cal M}(X)$ is the
quantum moduli of the heterotic string compactified on $Y$ are known as
Heterotic-type II dual pairs. Part of the beauty of these dual pairs is that
the $T$-duality on ${\cal M}(X)$ is inducing, when we read it in the variables
of the heterotic string compactified on $Y$, an $S$-duality transformation
on the dilaton field, that in fact is part of the non perturbative quantum
monodromy for the heterotic string on $Y$. Continuing with our translation
rules, the previous discussion can be summarized in the following diagram:

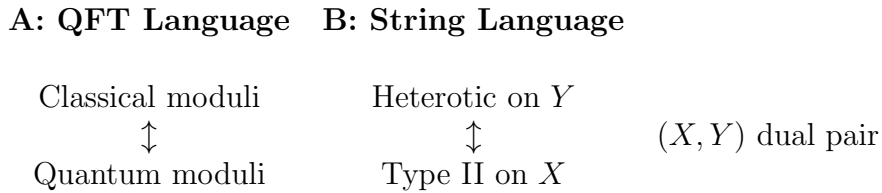
\vspace{4 mm}

\begin{center}
\begin{tabular}{cccc}

         &{\bf A: QFT Language} & {\bf B: String Language} & \\
         &                      &                       & \\
         & Classical moduli     & Heterotic on $Y$      & \\
         & $\updownarrow$       & $\updownarrow$        & $(X,Y)$ dual pair \\
         & Quantum moduli       & Type II on $X$        &
\end{tabular}
\end{center}

\vspace{4 mm}

To end this lectures, we wish just to mention a $N\!=\!2$ dual pair, defined
by heterotic compactification on $K3 \times T^{2}$ and type II on the
weighted projective space $W\IP^{(1,1,2,2,6)}$ \cite{kv}. This dual pair is the
natural candidate for recovering from the string the quantum moduli structure
of $N\!=\!2$ $SU(2)$ super Yang-Mills \cite{ow}, opening the possibility to
explore
how strings and gravitational effects modify the point particle limit, quantum
moduli physics described in these notes.

\vspace{8 mm}

\begin{center}
{\bf Acknowledgements.}
\end{center}

\vspace{4 mm}

It is pleasure for C. G. to thank M. J. Herrero, A. Dobado and F. Cornet for
their
invitation to give these lectures in such a lovely place. The work of C. G. is
supported in part by PB 92-1092, ERBCHRXCT920069 and OFES contract number
93.0083.
The work of R. H. is supported by a fellowship from UAM.

\newpage


\end{document}